\documentclass[epsfig,12pt]{article}
\usepackage{makeidx}
\usepackage{amsmath}
\usepackage{amsfonts}
\usepackage{amssymb}
\usepackage{graphicx}

\input epsf.sty
\newtheorem{theorem}{Theorem}
\newtheorem{acknowledgement}[theorem]{Acknowledgement}

\newtheorem{claim}[theorem]{Claim}

\makeatletter \@addtoreset{equation}{section}
\renewcommand{\theequation}{\thesection.\arabic{equation}}
\input{tcilatex}
\begin{document}

\title{\rightline{\mbox{\small
{LPHE-MS-11-03/CPM-11-03}}} \bigskip \textbf{On Dirac Zero Modes in
Hyperdiamond Model}}
\author{Lalla Btissam Drissi$^{a}$, El Hassan Saidi$^{a,b,c}$ \\
{\small a. INANOTECH-MAScIR, Institute of Nanomaterials and Nanotechnology,
Rabat, Morocco,}\\
{\small b. LPHE, Modelisation et Simulation, Facult\'{e} des Sciences Rabat,
Morocco}\\
{\small c. Centre of Physics and Mathematics, CPM-CNESTEN, Rabat-Morocco}}
\maketitle

\begin{abstract}
Using the $SU\left( 5\right) $ symmetry of the \emph{4D}\ hyperdiamond and
results on the study of \emph{4D} graphene given in \textrm{\cite{10}}, we
engineer a class of \emph{4D} lattice QCD fermions whose Dirac operators
have two zero modes. We show that generally the zero modes of the Dirac
operator in hyperdiamond fermions are captured by a tensor $\Omega _{\mu
}^{l}$ with $4\times 5$ complex components linking the Euclidean SO$\left(
4\right) $ vector $\mu ;$ and the 5-dimensional representation of $SU\left(
5\right) $. The \emph{Bori\c{c}i-Creutz} (BC) and the \emph{Karsten-Wilzeck}
(KW) models as well as their Dirac zero modes are rederived as particular
realizations of $\Omega _{\mu }^{l}$. Other features are also given.\newline
\textbf{Keywords}: Lattice QCD, \emph{Bori\c{c}i-Creutz} and \emph{%
Karsten-Wilzeck} models, \emph{4D} hyperdiamond, \emph{4D} graphene, $%
SU\left( 5\right) $ Symmetry.
\end{abstract}


\section{Introduction}

One of the highly important things in lattice \emph{QCD} is the need to have
a fermion action with a Dirac operator $\mathcal{D}$ having two zero modes
at the points $K$ and $K^{\prime }$ of the reciprocal space; so that they
could be interpreted as the two light quarks, up and down. In this regards, 
\emph{Michael Creutz} proposed few years ago an interesting \emph{4D}
lattice fermion action \textrm{\cite{1} }by extending the dispersion energy
relation of \emph{2D}\ graphene \textrm{\cite{2,3} to \emph{4D} }%
quaternions. The \emph{Creutz} fermions, which exhibit very useful
properties for dealing with lattice \emph{QCD}, was further developed by 
\emph{Artan} \emph{Bori\c{c}i} \textrm{\cite{4}} leading afterwards to the
so called \emph{Bori\c{c}i-Creutz} (\emph{BC}) fermions \textrm{\cite{5,6}}.
This is a simple \emph{4D}\ fermion lattice model using a 4-component Dirac
spinor $\Psi _{\mathbf{r}}$ as well as a particular linear combination of
the $4\times 4$ matrices $\mathrm{\gamma }_{\mu }$. \emph{BC} lattice model
has too particularly chiral symmetry, preventing mass renormalization, and a
Dirac operator $\mathcal{D}_{BC}$ with two zero modes $K_{BC}$ and $%
K_{BC}^{\prime }$ that describe a degenerate doublet of quarks. The
engineering of these two zeros is mainly due to mimicking \emph{2D} graphene
as well as to a nice trick relying on using a "complexification" of the
Dirac matrices type $\Upsilon _{\mu }=\mathrm{\gamma }_{\mu }+i\mathrm{%
\gamma }_{\mu }^{\prime }$ and the remarkable linear combination $\mathrm{%
\gamma }_{1}+\mathrm{\gamma }_{2}+\mathrm{\gamma }_{3}+\mathrm{\gamma }_{4}$%
. This way of doing let suspect that the \emph{BC} model hides a more
fundamental property that controls, amongst others, the engineering of Dirac
operator zeros.

\ \ \ \ \newline
In this paper, we contribute to this matter by proposing a fermionic \emph{4D%
} lattice \emph{QCD}-like model based on $SU\left( 5\right) $ symmetry and
containing \emph{BC} and \emph{KW }\textrm{\cite{7,8}} fermions as
particular cases. In our proposal the \emph{4D} lattice, denoted as $\mathbb{%
L}_{4}$, is given by the hyperdiamond generated by the \emph{4} simple roots 
$\alpha _{1},$ $\alpha _{2},$ $\alpha _{3},$ $\alpha _{4}$ of $SU\left(
5\right) $ \textrm{\cite{9}}. Moreover, each site $\mathbf{r}_{A}$ of $%
\mathbb{L}_{4}$ has \emph{5} first nearest neighbors $\left( \mathbf{r}_{A}+a%
\mathbf{\lambda }_{l}\right) $ forming a 5-dimensional representation of the 
$SU\left( 5\right) $ symmetry. \newline
The proposed free massless fermion lattice action is given by 
\begin{equation}
\begin{tabular}{ll}
$\mathcal{S}$ $\sim $ & $\frac{i}{4a}\dsum\limits_{\mathbf{r}%
}\dsum\limits_{\mu =1}^{4}\left( \dsum\limits_{l=0}^{4}\bar{\Psi}_{\mathbf{r}%
}\left( \mathrm{\gamma }^{\mu }\mathrm{\Omega }_{\mu }^{l}\right) \Psi _{%
\mathbf{r+}a\mathbf{\lambda }_{l}}+\dsum\limits_{l=0}^{4}\bar{\Psi}_{\mathbf{%
r+}a\mathbf{\lambda }_{l}}\left( \mathrm{\gamma }^{\mu }\mathrm{\bar{\Omega}}%
_{\mu }^{l}\right) \Psi _{\mathbf{r}}\right) $%
\end{tabular}
\label{PR}
\end{equation}%
where $\Psi _{\mathbf{r}}$ lives on $\mathbb{L}_{4}$ and where the \emph{5}
vectors $\lambda _{0},$ $\lambda _{1},$ $\lambda _{2},$ $\lambda _{3},$ $%
\lambda _{4}$ stand for the weight vectors of the 5- dimensional
representation of $SU\left( 5\right) $ satisfying the basic property $%
\lambda _{0}+\lambda _{1}+\lambda _{2}+\lambda _{3}+\lambda _{4}=0$. This
identity has a physical interpretation in terms of conservation of total
momenta at each site $\mathbf{r}$; and has as well a strong link with the
linear combination $\mathrm{\gamma }_{1}+\mathrm{\gamma }_{2}+\mathrm{\gamma 
}_{3}+\mathrm{\gamma }_{4}-2\Gamma =0$ used in \emph{BC} model. The distance 
$a\left\Vert \mathbf{\lambda }_{l}\right\Vert =a\frac{2\sqrt{5}}{5}$ is the
parameter $d$ of the lattice. \newline
Notice that our lattice fermion model depends on the complex tensor $\mathrm{%
\Omega }_{\mu }^{l}$ whose features will be studied here in details. It is a
kind of a complex $4\times 5$ rectangular matrix that plays a quite similar
role to the matrix combinations $\mathrm{\gamma }_{\mu }+i\mathrm{\gamma }%
_{\mu }^{\prime }$, $\mathrm{\gamma }_{\mu }+i\mathrm{\gamma }_{4}\left(
1-\delta _{\mu 4}\right) $ and $\mathrm{\gamma }_{1}+\mathrm{\gamma }_{2}+%
\mathrm{\gamma }_{3}+\mathrm{\gamma }_{4}$ used in the \emph{BC} and \emph{KW%
} models \textrm{\cite{5,6}}. In the case of free \emph{BC} fermions, we
will show that the corresponding $\left( \mathrm{\Omega }_{\mu }^{l}\right)
_{\emph{BC}}$ tensor reads as%
\begin{equation}
\left( 
\begin{array}{ccccc}
2 & 1+2i & -1 & -1 & -1 \\ 
2 & -1 & 1+2i & -1 & -1 \\ 
2 & -1 & -1 & 1+2i & -1 \\ 
2 & -1 & -1 & -1 & 1+2i%
\end{array}%
\right) ,  \label{BC}
\end{equation}%
and for the \emph{KW} ones, the associated $\left( \mathrm{\Omega }_{\mu
}^{l}\right) _{\emph{KW}}$ is given by%
\begin{equation}
\left( 
\begin{array}{ccccc}
0 & i & 0 & 0 & 0 \\ 
0 & 0 & i & 0 & 0 \\ 
0 & 0 & 0 & i & 0 \\ 
2 & 1 & 1 & -2 & i-2%
\end{array}%
\right) .  \label{KW}
\end{equation}%
Notice also that, generally speaking, $\mathrm{\Omega }_{\mu }^{l}$
transforms in the bi-fundamental of $SO\left( 4\right) \times SU\left(
5\right) $; and should not be a tensor product; i.e $\mathrm{\Omega }_{\mu
}^{l}\neq \xi _{\mu }\times \zeta ^{l}$ in order to recover the continuum
limit. This $\mathrm{\Omega }_{\mu }^{l}$ links then the $SO\left( 4\right) $
and $SU\left( 5\right) $ fundamental representations respectively captured
by the indices $\mu =1,2,3,4$\ and $l=0,1,2,3,4$. It allows as well to
implement quark-quark-gluon interactions by letting it to depend on
positions $\mathbf{r}$. \newline
The tensor $\mathrm{\Omega }_{\mu }^{l}$ has another remarkable feature that
we are interested in here; it plays a crucial role in the engineering of
lattice Dirac operator with a generic number $N$ of zeros; including the
interesting case $N=2$. Splitting $\Omega _{\mu }^{l}$\ into real and
imaginary parts as $R_{\mu }^{l}+iJ_{\mu }^{l}$, it is clear that the set of
real parameters captured by the rectangular matrix $R_{\mu }^{l}$, and
similarly for $J_{\mu }^{l}$, is given by the real 20-dimensional moduli
space 
\begin{equation}
\mathfrak{R}_{20}=\frac{SO\left( 4,5\right) }{SO\left( 4\right) \times
SO\left( 5\right) },  \label{M}
\end{equation}%
so that the total moduli space of (\ref{PR}) is given by the
complexification of $\mathfrak{R}_{20}$ namely $\frac{U\left( 4,5\right) }{%
U\left( 4\right) \times U\left( 5\right) }$ with dimension $\dim _{R}U\left(
4,5\right) -\dim _{R}U\left( 4\right) -\dim _{R}U\left( 5\right) =40$. As
such, the zero modes of the lattice "Dirac" operator $\mathcal{D}_{k,\Omega
}\sim \gamma ^{\mu }\left( \Omega _{\mu }^{l}e^{ia\mathbf{k}.\mathbf{\lambda 
}_{l}}+\bar{\Omega}_{\mu }^{l}e^{-ia\mathbf{k}.\mathbf{\lambda }_{l}}\right) 
$ of the above lattice action (\ref{PR}) correspond to some particular
tensors $\left( \Omega _{\mu }^{l}\right) _{N\text{-zeros}}$; and so would
live on subspaces of $\mathfrak{R}_{20}\times \mathfrak{R}_{20}$. From this
moduli space view, operators $\mathcal{D}_{k,\Omega }$ that have two zero
modes have tensors type $\left( \Omega _{\mu }^{l}\right) _{2\text{-zeros}}$
which require the fixing $\left( 4+4\right) $ real moduli; and therefore
would live on real 32-dimensional subspaces $\mathfrak{M}_{32}$ of the
moduli subspace. Typical examples of $\mathfrak{M}_{32}$ is given by the
particular space%
\begin{equation}
\frac{U\left( 4,4\right) }{U\left( 4\right) \times U\left( 4\right) },
\end{equation}%
or more generally by $\mathfrak{R}_{20-n}\times \mathfrak{R}_{20-m}$ with $%
n+m=8$. In addition to the previous \emph{BC} and \emph{KW} examples with
two zeros, one can engineer others by solving appropriate constraint
relations or directly by making wise choices based using symmetries. This
analysis is developed in sub-section 4.2 from which we learn that the real $%
\left( R_{\mu }^{l}\right) _{BC}$ and imaginary $\left( J_{\mu }^{l}\right)
_{BC}$ parts of the \emph{BC} complex tensor $\left( \Omega _{\mu
}^{l}\right) _{\emph{BC}}$ should obey the constraint relations 
\begin{equation*}
\begin{tabular}{llll}
$\dsum\limits_{\nu =1}^{4}\left( R_{\mu }^{\nu }\right) _{BC}=-\left( R_{\mu
}^{0}\right) _{BC}$ & , & $\dsum\limits_{\nu =1}^{4}\left( J_{\mu }^{\nu
}\right) _{BC}=\left( R_{\mu }^{0}\right) _{BC}$ & ,%
\end{tabular}%
\end{equation*}%
Similar relations can be also written down for the tensors $\left( R_{\mu
}^{l}\right) _{KW}$ and $\left( J_{\mu }^{l}\right) _{KW}$ in the case of
the \emph{KW} model; see eqs (\ref{KWK}).\newline
The presentation of this paper is as follows: In section 2, we recall
briefly some results on the \emph{BC}\textrm{\ }and \emph{KW} models; in
particular the engineering of the two zeros of the corresponding Dirac
operator. We also give extensions of these models where the linear
combination $\Gamma _{5}=\mathrm{\gamma }_{1}+\mathrm{\gamma }_{2}+\mathrm{%
\gamma }_{3}+\mathrm{\gamma }_{4}$ is interpreted in terms of a fifth
direction; the four others are $\Gamma _{\mu }=\mathrm{\gamma }_{\mu }+i%
\mathrm{\gamma }_{\mu }^{\prime }$. These extensions may also be understood
as an indication for the existence of the tensor $\Omega _{\mu }^{l}$ that
captures data on the zeros of Dirac operators. In section 3, we develop our
proposal; and show, amongst others, that the Dirac operator following from (%
\ref{PR}) can also have two zero modes by making wise choices of $\mathrm{%
\Omega }_{\mu }^{l}$. In section 4, we re-derive the \emph{BC} and \emph{KW}
theories, and their extensions using $\Gamma _{5}$ term, from (\ref{PR}). In
section 5, we give a conclusion and a comment on gauged (\ref{PR}).

\section{\emph{BC/KW} fermions and beyond}

First we study the two zero modes $K_{BC}$ and $K_{BC}^{\prime }$ of the
Dirac operator $\mathcal{D}_{BC}$ of the \emph{BC} fermions; then we give
those $K_{KW}$ and $K_{KW}^{\prime }$ of the Dirac operator $\mathcal{D}%
_{KW} $ of \emph{KW} model. \newline
We also use this section to give an extension of \emph{BC} and \emph{KW}
models where the term $\bar{\Psi}_{\mathbf{r}}\Gamma \Psi _{\mathbf{r}}$ is
modified as $\bar{\Psi}_{\mathbf{r}}\Gamma ^{5}\Psi _{\mathbf{r}+a\mathbf{5}%
} $.\newline
The analysis given in this section is useful to fix the ideas; but also for
later use since \emph{BC,} \emph{KW} models and their extensions can be
re-derived from the lattice action (\ref{PR}) by using wise choices of the
tensor $\mathrm{\Omega }_{\mu }^{l}$.

\subsection{\emph{BC} fermions: the free model}

Following \textrm{\cite{5,6}} and using the 4-component Dirac spinors $\Psi
_{\mathbf{r}}=\left( \phi _{\mathbf{r}}^{a},\bar{\chi}_{\mathbf{r}}^{\dot{a}%
}\right) $, the lattice action of \emph{free} \emph{BC} fermions reads in
the position space, by dropping mass term $m_{0}$, as follows:%
\begin{equation}
\begin{tabular}{lll}
$\mathcal{S}_{BC}$ $\sim $ & $\frac{1}{2a}\dsum\limits_{\mathbf{r}}\left(
\dsum\limits_{\mu =1}^{4}\bar{\Psi}_{\mathbf{r}}\Upsilon ^{\mu }\Psi _{%
\mathbf{r+}a\mathbf{\mu }}-\dsum\limits_{\mu =1}^{4}\bar{\Psi}_{\mathbf{r}+a%
\mathbf{\mathbf{\mu }}}\bar{\Upsilon}^{\mu }\Psi _{\mathbf{r}}\right) -\frac{%
2i}{a}\dsum\limits_{\mathbf{r}}\bar{\Psi}_{\mathbf{r}}\Gamma \Psi _{\mathbf{r%
}}$ &  \\ 
&  & 
\end{tabular}
\label{BBC}
\end{equation}%
where, for simplicity, we have dropped out gauge interactions; and where and 
$\Upsilon ^{\mu }=\mathrm{\gamma }^{\mu }+i\mathrm{\gamma }^{\prime \mu }$;
which is a kind of complexification of the Dirac matrices. \newline
Moreover, the matrix $\Gamma $ appearing in the last term is a $4\times 4$
matrix linked to $\mathrm{\gamma }^{\mu }$, $\mathrm{\gamma }^{\prime \mu }$
as follows: 
\begin{equation}
\begin{tabular}{lllll}
$\mathrm{\gamma }^{\prime \mu }$ & $=\Gamma -\mathrm{\gamma }^{\mu }$ & , & $%
\mathrm{\gamma }^{\mu }\mathrm{\gamma }^{\nu }+\mathrm{\gamma }^{\nu }%
\mathrm{\gamma }^{\mu }$ & $=2\delta ^{\mu \nu }$ \\ 
$2\Gamma $ & $=\dsum\limits_{\mu =1}^{4}\mathrm{\gamma }^{\mu }$ & , & $%
\mathrm{\gamma }^{\mu }+i\mathrm{\gamma }^{\prime \mu }$ & $=\Upsilon ^{\mu
} $%
\end{tabular}
\label{AG}
\end{equation}%
Mapping (\ref{BBC}) to the reciprocal space, we have 
\begin{equation}
\begin{tabular}{lll}
$\mathcal{S}_{BC}$ $\sim $ & $\dsum\limits_{\mathbf{k}}\bar{\Psi}_{\mathbf{k}%
}\mathcal{D}_{BC}\Psi _{\mathbf{k}}$ & 
\end{tabular}
\label{DC}
\end{equation}%
where the massless Dirac operator $\mathcal{D}_{BC}$ is given by 
\begin{equation}
\begin{tabular}{lll}
$\mathcal{D}_{BC}=$ & $+\frac{1}{2a}\left( \Upsilon _{\mu }-\bar{\Upsilon}%
_{\mu }\right) \cos \left( ak_{\mu }\right) $ &  \\ 
&  &  \\ 
& $+\frac{i}{2a}\left( \Upsilon _{\mu }+\bar{\Upsilon}_{\mu }\right) \sin
\left( ak_{\mu }\right) $ $-\frac{2i}{a}\Gamma $ & . \\ 
&  & 
\end{tabular}
\label{TIT}
\end{equation}%
Upon using $\Upsilon _{\mu }+\bar{\Upsilon}_{\mu }=2\mathrm{\gamma }_{\mu }$
and $\Upsilon _{\mu }-\bar{\Upsilon}_{\mu }=2i\mathrm{\gamma }_{\mu
}^{\prime }$, we can be put $\mathcal{D}_{BC}$ in the form 
\begin{equation}
\mathcal{D}_{BC}=D_{\mathbf{k}}+\bar{D}_{\mathbf{k}}-\frac{2i}{a}\Gamma
\label{DIR}
\end{equation}%
with 
\begin{equation}
\begin{tabular}{llll}
$D_{\mathbf{k}}=\frac{i}{a}\left( \dsum\limits_{\mu =1}^{4}\mathrm{\gamma }%
^{\mu }\sin ak_{\mu }\right) $ & , & $\bar{D}_{\mathbf{k}}=\frac{i}{a}\left(
\dsum\limits_{\mu =1}^{4}\mathrm{\gamma }^{\prime \mu }\cos ak_{\mu }\right) 
$ & , \\ 
&  &  & 
\end{tabular}
\label{IR}
\end{equation}%
where $k_{\mu }=\mathbf{k}.\mathbf{\mu }$. In the next subsection and in
section 5, we will derive the explicit expression of these $k_{\mu }$'s in
terms of the weight vectors $\mathbf{\lambda }_{l}$ of the 5-dimensional
representation of the $SU\left( 5\right) $ symmetry as well as useful
relations. \newline
The zero modes of $\mathcal{D}_{BC}$ are points in the reciprocal space;
they are obtained by solving $\mathcal{D}_{BC}=0$; which leads to the
following condition 
\begin{equation}
\begin{tabular}{ll}
$\dsum\limits_{\mu =1}^{4}\mathrm{\gamma }^{\mu }\left( \sin aK_{\mu }-\cos
aK_{\mu }\right) -\Gamma \left( 2-\dsum\limits_{\mu =1}^{4}\cos aK_{\mu
}\right) =0$ & .%
\end{tabular}%
\end{equation}%
This condition is a constraint relation on the wave vector components $%
K_{\mu }$; it is solved by the two following wave vectors:%
\begin{equation}
\begin{tabular}{llll}
point $K_{BC}$ & : & $K_{1}=K_{2}=K_{3}=K_{4}=0$ & , \\ 
point $K_{BC}^{\prime }$ & : & $K_{1}^{\prime }=K_{2}^{\prime
}=K_{3}^{\prime }=K_{4}^{\prime }=\frac{\pi }{2a}$ & ,%
\end{tabular}
\label{BCM}
\end{equation}%
and\ are interpreted in lattice \emph{QCD} as associated with the light
quarks up and down. \newline
Notice that if giving up the $\mathrm{\gamma }_{\mu }^{\prime }$- terms in
eqs(\ref{BBC}-\ref{DC}); i.e $\mathrm{\gamma }_{\mu }^{\prime }\rightarrow 0$%
, the remaining terms in $\mathcal{D}_{BC}$ namely $D_{\mathbf{k}}\sim 
\mathrm{\gamma }^{\mu }\sin aK_{\mu }$ have \emph{16} zero modes given by
the wave components $K_{\mu }=0,\pi $. By switching on the $\mathrm{\gamma }%
_{\mu }^{\prime }$-terms, \emph{14} zeros are removed.

\subsection{Beyond \emph{BC} model}

Although very important, the lattice action $\mathcal{S}_{BC}$ is somehow
very particular; it let suspecting to hide a more fundamental property which
can be explicitly exhibited by using symmetries. This will be developed in
details in next sections; but in due time, notice the three following
features that can be understood as a way to exhibit the $SU\left( 5\right) $
symmetry in \emph{BC} model.

\begin{description}
\item[1)] The price to pay for getting a Dirac operator with two zero modes
is the involvement of the complexified Dirac matrices $\Upsilon ^{\mu },$ $%
\bar{\Upsilon}^{\mu }$ as well as the particular matrix $\Gamma $. Despite
that it breaks the $SO\left( 4\right) $ Lorentz symmetry since it can be
written as 
\begin{equation}
\begin{tabular}{ll}
$\Gamma =\frac{1}{2}\dsum\limits_{\mu =1}^{4}\gamma ^{\mu }\upsilon _{\mu }$
& 
\end{tabular}%
\end{equation}%
with%
\begin{equation}
\upsilon _{\mu }=\left( 
\begin{array}{c}
1 \\ 
1 \\ 
1 \\ 
1%
\end{array}%
\right)  \label{VM}
\end{equation}%
violating explicitly the SO$\left( 4\right) $, the matrix $\Gamma $ plays an
important role in studying the zero modes. The expression of the matrix $%
\Gamma $ (\ref{AG}) should be thought of as associated precisely with the
solution of the constraint relation%
\begin{equation}
2\Gamma -\dsum\limits_{\mu =1}^{4}\mathrm{\gamma }^{\mu }=0
\end{equation}%
that is required by a hidden symmetry of the \emph{BC} model namely the SU$%
\left( 5\right) $ symmetry of the \emph{4D} hyperdiamond; see also the
correspondence given in section 4 of ref \textrm{\cite{10}}.

\item[2)] the \emph{BC}\ action $\mathcal{S}_{BC}$ lives on a \emph{4D}
lattice $\mathbb{L}_{4}^{BC}$ generated by $\mathbf{\mu \equiv v}_{\mu }$;
i.e the vectors 
\begin{equation}
\begin{tabular}{lllll}
$\mathbf{v}_{1}\mathbf{=}\left( 
\begin{array}{c}
v_{1}^{x} \\ 
v_{1}^{y} \\ 
v_{1}^{z} \\ 
v_{1}^{t}%
\end{array}%
\right) ,$ & $\mathbf{v}_{2}\mathbf{=}\left( 
\begin{array}{c}
v_{2}^{x} \\ 
v_{2}^{y} \\ 
v_{2}^{z} \\ 
v_{2}^{t}%
\end{array}%
\right) ,$ & $\mathbf{v}_{3}\mathbf{=}\left( 
\begin{array}{c}
v_{3}^{x} \\ 
v_{3}^{y} \\ 
v_{3}^{z} \\ 
v_{3}^{t}%
\end{array}%
\right) ,$ & $\mathbf{v}_{4}\mathbf{=}\left( 
\begin{array}{c}
v_{4}^{x} \\ 
v_{4}^{y} \\ 
v_{4}^{z} \\ 
v_{4}^{t}%
\end{array}%
\right) $ &  \\ 
&  &  &  & 
\end{tabular}%
\end{equation}%
These \ $\mathbf{\mu }$-vectors look somehow ambiguous to interpret by using
the analogy with \emph{4D} graphene prototype; they are also ambiguous to
interpret from the SU$\left( 5\right) $ symmetry view.\ Indeed, to each site 
\begin{equation}
\mathbf{r}\in \mathbb{L}_{4}^{BC},  \notag
\end{equation}%
there should be \emph{5} first nearest neighbors that are rotated by $%
SU\left( 5\right) $ symmetry. But from the \emph{BC} action we learn that
the first nearest neighbors to each site $\mathbf{r}$ are:%
\begin{equation}
\begin{tabular}{lll}
$\mathbf{r}$ & $\rightarrow $ & $\left\{ 
\begin{array}{c}
\mathbf{r+}a\mathbf{v}_{1} \\ 
\mathbf{r+}a\mathbf{v}_{2} \\ 
\mathbf{r+}a\mathbf{v}_{3} \\ 
\mathbf{r+}a\mathbf{v}_{4}%
\end{array}%
\right. $,%
\end{tabular}%
\end{equation}%
The fifth missing one, namely%
\begin{equation}
\begin{tabular}{llll}
$\mathbf{r}$ & $\rightarrow $ & $\mathbf{r+}a\mathbf{v}_{5}$ & ,%
\end{tabular}%
\end{equation}%
may be interpreted in the \emph{BC} fermions as associated with the extra
term involving the matrix $\Gamma $.

\item[3)] To take into account the five nearest neighbors, we then have to
use the rigorous correspondence%
\begin{equation}
\begin{tabular}{llll}
$\Gamma ^{\mu }\rightarrow \mathbf{v}_{\mu }\mathbf{,}$ & and & $\Gamma
^{5}\rightarrow \mathbf{v}_{5}$ & 
\end{tabular}%
\end{equation}%
which can be also written in a combined form as follows%
\begin{equation}
\begin{tabular}{ll}
$\Gamma ^{M}\rightarrow \mathbf{v}_{M}$ & $\mathbf{,}$%
\end{tabular}%
\end{equation}%
with%
\begin{equation}
\begin{tabular}{lll}
$\Gamma ^{M}=\left( \Gamma ^{\mu },\Gamma ^{5}\right) $ & and & $\mathbf{v}%
_{M}=\left( \mathbf{v}_{\mu },\mathbf{v}_{5}\right) .$%
\end{tabular}%
\end{equation}%
Because of the $SU\left( 5\right) $ symmetry properties \textrm{\cite{9}},
we also have to require the condition%
\begin{equation}
\begin{tabular}{ll}
$\mathbf{v}_{1}+\mathbf{\mathbf{v}_{2}+\mathbf{v}_{3}+\mathbf{v}_{4}+v}%
_{5}=0 $ & ,%
\end{tabular}%
\end{equation}%
characterizing the \emph{5} first nearest neighbors. \newline
To determine the explicit expressions of the matrices $\Gamma _{M}$ in terms
of the usual Dirac ones, we modify the \emph{BC} action (\ref{BBC}) as
follows%
\begin{equation}
\begin{tabular}{lll}
$\mathcal{S}_{BC}^{\prime }$ $\sim $ & $\frac{1}{2a}\dsum\limits_{\mathbf{r}%
}\left( \dsum\limits_{M=1}^{5}\bar{\Psi}_{\mathbf{r}}\Gamma ^{M}\Psi _{%
\mathbf{r+}a\mathbf{v}_{M}}-\dsum\limits_{M=1}^{5}\bar{\Psi}_{\mathbf{r+av}%
_{M}}\Gamma ^{M}\Psi _{\mathbf{r}}\right) $ & , \\ 
&  & 
\end{tabular}
\label{EXT}
\end{equation}%
exhibiting both $SO\left( 4\right) $ and SU$\left( 5\right) $ symmetries and
leading to the following free Dirac operator%
\begin{equation}
\begin{tabular}{lll}
$\mathcal{D}=$ & $\frac{i}{2a}\dsum\limits_{\mu =1}^{4}\left( \Gamma _{\mu }+%
\bar{\Gamma}_{\mu }\right) \sin \left( ak_{\mu }\right) +\frac{i}{2a}\left(
\Gamma _{5}+\bar{\Gamma}_{5}\right) \sin \left( ak_{5}\right) $ &  \\ 
& $\frac{1}{2a}\dsum\limits_{\mu =1}^{4}\left( \Gamma _{\mu }-\bar{\Gamma}%
_{\mu }\right) \cos \left( ak_{\mu }\right) +\frac{1}{2a}\left( \Gamma _{5}-%
\bar{\Gamma}_{5}\right) \cos \left( ak_{5}\right) $ & 
\end{tabular}
\label{EG}
\end{equation}%
where $k_{M}=\mathbf{k.v}_{M}$ and where 
\begin{equation}
\begin{tabular}{llll}
$\dprod\limits_{M=1}^{5}e^{iak_{M}}=1$ & , & $\dsum\limits_{M=1}^{5}k_{M}=0$
& ,%
\end{tabular}%
\end{equation}%
expressing the conservation of total momenta at each lattice site.\newline
Equating with (\ref{TIT}-\ref{DIR}-\ref{IR}), we get the identities%
\begin{equation}
\begin{tabular}{ll}
$\Upsilon _{\mu }+\bar{\Upsilon}_{\mu }=\Gamma _{\mu }+\bar{\Gamma}_{\mu }$
& , \\ 
$\Upsilon _{\mu }-\bar{\Upsilon}_{\mu }=\Gamma _{\mu }-\bar{\Gamma}_{\mu }$
& ,%
\end{tabular}
\label{Ga}
\end{equation}%
and 
\begin{equation}
\begin{tabular}{ll}
$\frac{i}{2a}\left( \Gamma _{5}+\bar{\Gamma}_{5}\right) \sin \left(
ak_{5}\right) +\frac{1}{2a}\left( \Gamma _{5}-\bar{\Gamma}_{5}\right) \cos
\left( ak_{5}\right) =-\frac{4i}{2a}\Gamma $ & . \\ 
& 
\end{tabular}%
\end{equation}%
Eqs(\ref{Ga}) are solved by $\Gamma _{\mu }=\Upsilon _{\mu }$; that is%
\begin{equation}
\Gamma _{\mu }=\mathrm{\gamma }^{\mu }+i\mathrm{\gamma }^{\prime \mu }=%
\mathrm{\gamma }^{\mu }+i\left( \Gamma -\mathrm{\gamma }^{\mu }\right)
\end{equation}%
while 
\begin{equation}
\begin{tabular}{llll}
$\Gamma _{5}=-2i\Gamma $ & for & $\sin \left( ak_{5}\right) =0$ & , \\ 
&  &  &  \\ 
$\Gamma _{5}=-2\Gamma $ & for & $\sin \left( ak_{5}\right) =1$ & . \\ 
&  &  & 
\end{tabular}%
\end{equation}%
where $k_{5}=-\left( k_{1}+k_{2}+k_{3}+k_{4}\right) $.
\end{description}

\ \ \newline
In this 5-dimensional approach, to be further be developed in next sections,
the ambiguity in dealing with the $\mathbf{\mu }$-vectors is overcome; and
the underlying $SO\left( 4\right) $ and $SU\left( 5\right) $ symmetries of
the model in reciprocal space are explicitly exhibited.

\subsection{KW fermions}

The \emph{BC} fermions described above is not the only model that has a
Dirac operator with two zero modes. There is an other remarkable model that
has been considered in recent literature \textrm{\cite{5,6}}. This is given
by the so called \emph{Karsten-Wilzeck} model \textrm{\cite{7,8}} whose free
fermion lattice action reads as follows,%
\begin{equation}
\begin{tabular}{lll}
$\mathcal{S}_{KW}$ $\sim $ & $\frac{1}{2a}\dsum\limits_{\mathbf{r}}\left(
\dsum\limits_{\mu =1}^{4}\bar{\Psi}_{\mathbf{r}}\left[ \mathrm{\gamma }^{\mu
}-i\mathrm{\gamma }_{4}\left( 1-\delta _{\mu 4}\right) \right] \Psi _{%
\mathbf{r+}a\mathbf{\mu }}\right) +\dsum\limits_{\mathbf{r}}\bar{\Psi}_{%
\mathbf{r}}\left[ m_{0}+\frac{3i}{a}\mathrm{\gamma }_{4}\right] \Psi _{%
\mathbf{r}}$ &  \\ 
& $-\frac{1}{2a}\dsum\limits_{\mathbf{r}}\left( \dsum\limits_{\mu =1}^{4}%
\bar{\Psi}_{\mathbf{r+a\mathbf{\mu }}}\left[ \mathrm{\gamma }^{\mu }+i%
\mathrm{\gamma }_{4}\left( 1-\delta _{\mu 4}\right) \right] \Psi _{\mathbf{r}%
}\right) .$ & 
\end{tabular}%
\end{equation}%
This is quite similar to (\ref{BBC}); it is obtained by making the following
substitutions 
\begin{equation}
\begin{tabular}{lllll}
$\mathrm{\gamma }_{\mu }^{\prime }\longrightarrow \mathrm{\gamma }_{4}\left(
1-\delta _{\mu 4}\right) $ & , & $\Gamma \longrightarrow $ & $\frac{3i}{a}%
\mathrm{\gamma }_{4}$ & .%
\end{tabular}
\label{m}
\end{equation}%
Moreover mapping the above lattice action to the reciprocal space, the Dirac
operator $\mathcal{D}_{KW}$ reads as, upon dropping the bare mass $m_{0}$,
as follows:%
\begin{equation}
\begin{tabular}{lll}
$\mathcal{D}_{KW}$ & $=$ & $\dsum\limits_{\mu =1}^{4}\mathrm{\gamma }^{\mu
}\sin ak_{\mu }+\frac{i}{a}\mathrm{\gamma }^{4}\dsum\limits_{\mu
=1}^{3}\left( 1-\cos ak_{\mu }\right) =0.$%
\end{tabular}
\label{n}
\end{equation}%
This operator has also two zero modes located at the points $K_{KW}$ and $%
K_{KW}^{\prime }$,%
\begin{equation}
\begin{tabular}{lll}
$K_{KW}$ & $=\left( 0,0,0,0\right) $ &  \\ 
$K_{KW}^{\prime }$ & $=\left( 0,0,0,\frac{\pi }{a}\right) $ & 
\end{tabular}
\label{ZE}
\end{equation}%
The comments we gave in the case of \emph{BC }fermions apply as well to this
case. In particular, we have the \emph{extended}-\emph{KW} action 
\begin{equation}
\begin{tabular}{lll}
$\mathcal{S}_{KW}^{\prime }$ $\sim $ & $\frac{1}{2a}\dsum\limits_{\mathbf{r}%
}\left( \dsum\limits_{M=1}^{5}\bar{\Psi}_{\mathbf{r}}\Gamma ^{M}\Psi _{%
\mathbf{r+}a\mathbf{v}_{M}}-\dsum\limits_{M=1}^{5}\bar{\Psi}_{\mathbf{r+av}%
_{M}}\Gamma ^{M}\Psi _{\mathbf{r}}\right) $ & ,%
\end{tabular}
\label{XE}
\end{equation}%
where now 
\begin{equation}
\begin{tabular}{llll}
$\Gamma _{\mu }$ & \multicolumn{2}{l}{$=\mathrm{\gamma }^{\mu }+i\mathrm{%
\gamma }_{4}\left( 1-\delta _{\mu 4}\right) $} & , \\ 
$\Gamma _{5}$ & \multicolumn{2}{l}{$=-\frac{6i}{a}\mathrm{\gamma }_{4},$ \
for $\sin \left( ak_{5}\right) =0$} & , \\ 
$\Gamma _{5}$ & \multicolumn{2}{l}{$=-\frac{6}{a}\mathrm{\gamma }_{4}$, \ \
for $\sin \left( ak_{5}\right) =1$} & . \\ 
&  &  & 
\end{tabular}%
\end{equation}%
In section 5, we will show that the \emph{BC} and \emph{KW} fermions as well
as their extended versions with lattice actions $\mathcal{S}_{BC}^{\prime }$
and $\mathcal{S}_{KW}^{\prime }$ respectively given by (\ref{EXT})-(\ref{XE}%
), correspond in fact to particular realizations of the tensor $\Omega _{\mu
}^{l}$ of the proposal (\ref{PR}).

\section{Hyperdiamond fermions}

Before describing the proposal (\ref{PR}), it is interesting to start by
recalling useful features of the hyperdiamond $\mathbb{L}_{4}$\textrm{\ \cite%
{9,10,11}}.

\subsection{the \emph{4D} lattice}

The lattice $\mathbb{L}_{4}$ is a \emph{4D} lattice made by the
superposition of two sublattices $\mathcal{A}$ and $\mathcal{B}$ with
respective coordinates positions,%
\begin{equation}
\begin{tabular}{lllll}
$\frac{\mathbf{r}_{A}}{a}=$ & \multicolumn{3}{l}{$N_{1}\mathbf{\alpha }%
_{1}+N_{2}\mathbf{\alpha }_{2}+N_{3}\mathbf{\alpha }_{3}+N_{4}\mathbf{\alpha 
}_{4}$} & , \\ 
$\mathbf{r}_{B}=$ & $\mathbf{r}_{A}+d\frac{\sqrt{5}}{2}\mathbf{\lambda }_{l}$
& , & $a=d\frac{\sqrt{5}}{2}$ & ,%
\end{tabular}
\label{RB}
\end{equation}%
where the $N_{i}$'s are arbitrary integers and the $\mathbf{\lambda }_{l}$'s
are the weight vectors of the 5-dimensional representation of $SU\left(
5\right) $. \textrm{For a related construction see} \textrm{\cite{15,16}}.%
\newline
Recall that each $\mathbf{\lambda }_{l}$ weight has 4 components%
\begin{equation}
\mathbf{\lambda }_{l}^{\mu }=\left( \mathbf{\lambda }_{l}^{1},\mathbf{%
\lambda }_{l}^{2},\mathbf{\lambda }_{l}^{3},\mathbf{\lambda }_{l}^{4}\right)
,\qquad l=0,1,2,3,4,  \label{US}
\end{equation}%
obeying the group representation identities 
\begin{equation}
\begin{tabular}{llll}
$\lambda _{0}^{\mu }+\lambda _{1}^{\mu }+\lambda _{2}^{\mu }+\lambda
_{3}^{\mu }+\lambda _{4}^{\mu }=0$ & , & $\mu =1,...,4$ & .%
\end{tabular}
\label{SU}
\end{equation}%
These identities can be interpreted physically in terms of the conservation
of total momenta at each site of $\mathbb{L}_{4}$.%
\begin{equation}
\begin{tabular}{ll}
$\dprod\limits_{l=0}^{4}\exp \left( \frac{ia}{\hbar }\dsum\limits_{\mu
=1}^{4}p_{\mu }\mathbf{\lambda }_{l}^{\mu }\right) =\dprod\limits_{\mu
=1}^{4}\exp \left( \frac{ia}{\hbar }\dsum\limits_{l=0}^{4}p_{\mu }\mathbf{%
\lambda }_{l}^{\mu }\right) =1$ & .%
\end{tabular}%
\end{equation}%
Notice that the components $\mathbf{\lambda }_{l}^{\mu }$ can be expressed
in various, but equivalent, ways. Explicit ways were used in \textrm{\cite%
{9,10,11}}; see also \textrm{\cite{12}}, but, to our understanding, the more
convenient and significant way to do is the one using the $SU\left( 5\right) 
$ simple roots as follows: 
\begin{equation}
\begin{tabular}{ll}
$\mathbf{\lambda }_{0}=+\frac{4}{5}\mathbf{\alpha }_{1}+\frac{3}{5}\mathbf{%
\alpha }_{2}+\frac{2}{5}\mathbf{\alpha }_{3}+\frac{1}{5}\mathbf{\alpha }%
_{4}\ $ & , \\ 
$\mathbf{\lambda }_{1}=-\frac{1}{5}\mathbf{\alpha }_{1}+\frac{3}{5}\mathbf{%
\alpha }_{2}+\frac{2}{5}\mathbf{\alpha }_{3}+\frac{1}{5}\mathbf{\alpha }%
_{4}\ $ & , \\ 
$\mathbf{\lambda }_{2}=-\frac{1}{5}\mathbf{\alpha }_{1}-\frac{2}{5}\mathbf{%
\alpha }_{2}+\frac{2}{5}\mathbf{\alpha }_{3}+\frac{1}{5}\mathbf{\alpha }%
_{4}\ $ & , \\ 
$\mathbf{\lambda }_{3}=-\frac{1}{5}\mathbf{\alpha }_{1}-\frac{2}{5}\mathbf{%
\alpha }_{2}-\frac{3}{5}\mathbf{\alpha }_{3}+\frac{1}{5}\mathbf{\alpha }%
_{4}\ $ & , \\ 
$\mathbf{\lambda }_{4}=-\frac{1}{5}\mathbf{\alpha }_{1}-\frac{2}{5}\mathbf{%
\alpha }_{2}-\frac{3}{5}\mathbf{\alpha }_{3}-\frac{4}{5}\mathbf{\alpha }%
_{4}\ $ & . \\ 
& 
\end{tabular}
\label{AL}
\end{equation}%
These vectors capture the fact that each site $\mathbf{r}_{A}$ in the
sublattice $\mathcal{A}$ has \emph{5} first nearest neighbors in the
sublattice $\mathcal{B}$,%
\begin{equation}
\begin{tabular}{ll|ll}
{\small one site} & \multicolumn{2}{l}{$\rightarrow $} & \emph{5} {\small %
first nearest neighbors} \\ \hline
&  &  &  \\ 
$\mathbf{r}_{A}\in \mathcal{A}$ &  &  & $\left\{ 
\begin{array}{c}
\mathbf{r}_{A}+a\mathbf{\lambda }_{0} \\ 
\mathbf{r}_{A}+a\mathbf{\lambda }_{1} \\ 
\mathbf{r}_{A}+a\mathbf{\lambda }_{2} \\ 
\mathbf{r}_{A}+a\mathbf{\lambda }_{3} \\ 
\mathbf{r}_{A}+a\mathbf{\lambda }_{4}%
\end{array}%
\right. \in \mathcal{B}$ \\ 
&  &  &  \\ \hline
\end{tabular}%
\end{equation}%
Notice also that the expression of the $\mathbf{\lambda }_{l}$'s in terms of
the roots is helpful not only for calculations; but also for exhibiting
manifestly the role of the $SU\left( 5\right) $ symmetry.\newline
For later use, we also need the dual vectors,%
\begin{equation}
\mathbf{G}=\frac{2\pi N_{1}}{a}\mathbf{\omega }_{1}+\frac{2\pi N_{2}}{a}%
\mathbf{\omega }_{2}+\frac{2\pi N_{3}}{a}\mathbf{\omega }_{3}+\frac{2\pi
N_{4}}{a}\mathbf{\omega }_{4},  \label{G}
\end{equation}%
obeying the reciprocal lattice property%
\begin{equation}
\exp \left( ia\mathbf{G.r}_{A}\right) =1,  \label{G1}
\end{equation}%
thanks to the duality relation $\mathbf{\omega }_{i}.\mathbf{\alpha }%
_{j}=\delta _{ij}$.

\subsection{the free model}

The proposal (\ref{PR}) involves, in addition to the Dirac spinor $\Psi _{%
\mathbf{r}}$ and the $\mathrm{\gamma }_{\mu }$ matrices, the remarkable
complex tensor $\mathrm{\Omega }_{\mu }^{l}$; this is a bi-fundamental
representation of $SO\left( 4\right) \times SU\left( 5\right) $ given by the
complex $4\times 5$ matrix,%
\begin{equation}
\Omega _{\mu }^{l}=\left( 
\begin{array}{ccccc}
\Omega _{1}^{0} & \Omega _{1}^{1} & \Omega _{1}^{2} & \Omega _{1}^{3} & 
\Omega _{1}^{4} \\ 
\Omega _{2}^{0} & \Omega _{2}^{1} & \Omega _{2}^{2} & \Omega _{2}^{3} & 
\Omega _{2}^{4} \\ 
\Omega _{3}^{0} & \Omega _{3}^{1} & \Omega _{3}^{2} & \Omega _{3}^{3} & 
\Omega _{3}^{4} \\ 
\Omega _{4}^{0} & \Omega _{4}^{1} & \Omega _{4}^{2} & \Omega _{4}^{3} & 
\Omega _{4}^{4}%
\end{array}%
\right) ,
\end{equation}%
that allows to link the lattice Euclidean space time index $\mu $, capturing
a 4-component vector of $SO\left( 4\right) $; and the index $l$ of the
5-dimensional representation of the $SU\left( 5\right) $ symmetry of the
hyperdiamond. In some sense, it plays the role of $\mathrm{\gamma }_{\mu
}^{\prime }$ and $\Gamma $ in the \emph{BC}\ model.\newline
Notice that the action (\ref{PR}) can be also written in a similar form (\ref%
{EXT}), as follows 
\begin{equation}
\begin{tabular}{ll}
$\mathcal{S}$ $\sim $ & $\frac{i}{4a}\dsum\limits_{\mathbf{r}}\left(
\dsum\limits_{l=0}^{4}\bar{\Psi}_{\mathbf{r}}\Gamma ^{l}\Psi _{\mathbf{r+}a%
\mathbf{\lambda }_{l}}+\dsum\limits_{l=0}^{4}\bar{\Psi}_{\mathbf{r+}a\mathbf{%
\lambda }_{l}}\bar{\Gamma}^{l}\Psi _{\mathbf{r}}\right) $%
\end{tabular}%
\end{equation}%
where now the $\Gamma ^{l}$ has the following general form%
\begin{equation}
\begin{tabular}{llll}
$\Gamma ^{l}=\left( \dsum\limits_{\mu =1}^{4}\mathrm{\gamma }^{\mu }\mathrm{%
\Omega }_{\mu }^{l}\right) $ & , & $\bar{\Gamma}^{l}=\left(
\dsum\limits_{\mu =1}^{4}\mathrm{\gamma }^{\mu }\mathrm{\bar{\Omega}}_{\mu
}^{l}\right) $ & .%
\end{tabular}%
\end{equation}%
In the reciprocal space, our massless free fermion lattice action reads as%
\begin{equation}
\begin{tabular}{ll}
$\mathcal{S}$ $\sim $ & $\dsum\limits_{\mathbf{k}}\left( \dsum\limits_{\mu
=1}^{4}\bar{\Psi}_{\mathbf{k}}\mathcal{D}\Psi _{\mathbf{k}}\right) $%
\end{tabular}
\label{RP}
\end{equation}%
with Dirac operator given by 
\begin{equation}
\mathcal{D}=\frac{i}{4a}\mathrm{\gamma }^{\mu }\left( D_{\mu }+\bar{D}_{\mu
}\right) ,\qquad \frac{i}{4a}\left( D_{\mu }+\bar{D}_{\mu }\right) =\frac{1}{%
2}Tr\left( \mathrm{\gamma }_{\mu }\mathcal{D}\right) ,  \label{DI}
\end{equation}%
and where $D_{\mu }$ and its complex adjoint $\bar{D}_{\mu }$ read as
follows: 
\begin{equation}
\begin{tabular}{llll}
$D_{\mu }=\dsum\limits_{l=0}^{4}\mathrm{\Omega }_{\mu }^{l}e^{ia\mathbf{k}.%
\mathbf{\lambda }_{l}}$ & , & $\bar{D}_{\mu }=\dsum\limits_{l=0}^{4}\mathrm{%
\bar{\Omega}}_{\mu }^{l}e^{-ia\mathbf{k}.\mathbf{\lambda }_{l}}$ & .%
\end{tabular}
\label{3}
\end{equation}%
So the zero modes of the Dirac operator (\ref{DI}) are obtained by solving
the following constraint equation%
\begin{equation}
\begin{tabular}{ll}
$\func{Re}\left( \dsum\limits_{l=0}^{4}\mathrm{\Omega }_{\mu }^{l}e^{i\Phi
_{l}}\right) $ & $=0,$%
\end{tabular}
\label{16}
\end{equation}%
which reads explicitly like 
\begin{equation}
\begin{tabular}{ll}
$\dsum\limits_{l=0}^{4}\left[ \mathrm{\Omega }_{\mu }^{l}+\mathrm{\bar{\Omega%
}}_{\mu }^{l}\right] \cos \Phi _{l}+i\left[ \mathrm{\Omega }_{\mu }^{l}-%
\mathrm{\bar{\Omega}}_{\mu }^{l}\right] \sin \Phi _{l}$ & $=0$,%
\end{tabular}
\label{61}
\end{equation}%
where $\Phi _{l}$ are the relative phases of the wave functions $\Psi _{%
\mathbf{r}+a\mathbf{\lambda }_{l}}$ with respect to $\Psi _{\mathbf{r}}$;
they read \textrm{in the first Brillouin zone} as follows%
\begin{equation}
\begin{tabular}{llll}
$\Phi _{l}$ & $=a\left( \mathbf{k}.\mathbf{\lambda }_{l}\right) $ & , & $%
0\leq \Phi _{l}<2\pi $,%
\end{tabular}
\label{PH}
\end{equation}%
with%
\begin{equation}
\begin{tabular}{lll}
$\mathbf{k}.\mathbf{\lambda }_{l}$ & $=\dsum\limits_{\mu =1}^{4}k_{\mu
}.\lambda _{l}^{\mu }$ & .%
\end{tabular}
\label{KL}
\end{equation}%
to be denoted as $k_{l}\equiv \mathbf{k}.\mathbf{\lambda }_{l}$. Moreover,
because of the $SU\left( 5\right) $ symmetry (\ref{SU}), we also have the
constraint relation 
\begin{equation}
\dsum\limits_{i=0}^{4}\Phi _{l}=0,\qquad \func{mod}2\pi  \label{V}
\end{equation}%
showing that the \emph{5} phases $\Phi _{l}$ are related; so that one of
them; say $\Phi _{0}$, can be expressed in terms of the others like 
\begin{equation}
\Phi _{0}=-\left( \Phi _{1}+\Phi _{2}+\Phi _{3}+\Phi _{4}\right) .
\end{equation}%
Notice that if solving the constraint eqs(\ref{61}) as 
\begin{equation}
\Phi _{l}=\frac{2\pi N}{5},\qquad N=0,1,2,3,4,  \label{PA}
\end{equation}%
we end with a constraint relation on $\mathrm{\Omega }_{\mu }^{l}$ namely 
\begin{equation}
\sum_{l=0}^{5}\func{Re}\left( \mathrm{\Omega }_{\mu }^{l}\right) =0.
\label{p}
\end{equation}%
This condition can be solved by using eq(\ref{SU}); this shows that $\func{Re%
}\left( \mathrm{\Omega }_{\mu }^{l}\right) $ can be thought of as
proportional to the weight vectors $\lambda _{l}^{\mu }$ of SU$\left(
5\right) $. A particular realization of $\mathrm{\Omega }_{\mu }^{l}$ is%
\textrm{\ given by the one} considered in \textrm{\cite{9,10,11}}; and which
reads as follows%
\begin{equation}
\Omega _{\mu }^{l}=\left( 
\begin{array}{ccccc}
0 & \sqrt{5} & \sqrt{5} & -\sqrt{5} & -\sqrt{5} \\ 
0 & \sqrt{5} & -\sqrt{5} & -\sqrt{5} & \sqrt{5} \\ 
0 & \sqrt{5} & -\sqrt{5} & \sqrt{5} & -\sqrt{5} \\ 
-4 & 1 & 1 & 1 & 1%
\end{array}%
\right)  \label{r}
\end{equation}%
Notice in passing that the tensor $\left( \Omega _{\mu }^{l}\right) _{BC}$
and $\left( \Omega _{\mu }^{l}\right) _{KW}$ given in the introduction, and
which are associated with the \emph{BC} and \emph{KW} models, obey as well
the property (\ref{p}); this feature will be proved in next section. \newline
Below, we want to look for general solutions going beyond $\Phi _{l}=\frac{%
2\pi N}{5}$; and recover the special solutions (\ref{PA}) as particular
cases. This also allows us to get more information on the role of the tensor 
$\mathrm{\Omega }_{\mu }^{l}$.

\section{Solutions with two zero modes}

First we study a simple model with two zero modes having similar properties
as the \emph{BC} and \emph{KW} ones; then we analyze the generic case
involving the \emph{40} real moduli captured by the complex tensor $\Omega
_{\mu }^{l}$; and explore the link between the number of zero modes and the
moduli space of $\Omega _{\mu }^{l}$.

\subsection{Simple model}

This lattice model is a simple toy prototype having a lattice Dirac operator 
$\mathcal{D}\left( \mathbf{k}\right) $, of quite similar form to \emph{BC}
and \emph{KW} operators considered in section 2, sharing with eq(\ref{PR})
the three following features:

\begin{itemize}
\item the operator $\mathcal{D}\left( \mathbf{k}\right) $ has two zero modes
located at points $\mathbf{K=}\left( K_{1},K_{2},K_{3},K_{4}\right) $ and $%
\mathbf{K}^{\prime }\mathbf{=}\left( K_{1}^{\prime },K_{2}^{\prime
},K_{3}^{\prime },K_{4}^{\prime }\right) $ of the reciprocal space; that is: 
$\mathcal{D}\left( \mathbf{K}\right) =\mathcal{D}\left( \mathbf{K}^{\prime
}\right) =0,$

\item the continuum limit of $\mathcal{D}\left( \mathbf{k}\right) $ in the
neighborhood of the points $\mathbf{K}$ and $\mathbf{K}^{\prime }$ is given
by the usual free Dirac operator $\sum_{\mu }\gamma ^{\mu }k_{\mu }$ plus
higher order terms that have an interpretation in the Symanzik effective
theory.

\item it has the $SU\left( 5\right) $ symmetry of the hyperdiamond that
permits to mimick tight binding model of \emph{2D} graphene; there the
underlying \emph{2D} lattice is given by the honeycomb which is known to
have an $SU\left( 3\right) $ symmetry. The $SU\left( 5\right) $ encountered
in the present study is just the extension of the SU$\left( 3\right) $ to 
\emph{4D} dimensions.
\end{itemize}

\ \ \newline
The lattice action $\mathcal{S}$ of this model is of the form (\ref{PR});
but with a very particular tensor $\Omega _{\mu }^{l}$. It reads in the
reciprocal space like 
\begin{equation}
\mathcal{S}=\int \frac{d^{4}k}{\left( 2\pi \right) ^{4}}\bar{\Psi}\left( 
\mathbf{k}\right) \mathcal{D}\left( \mathbf{k}\right) \Psi \left( \mathbf{k}%
\right) .
\end{equation}%
For simplicity, we often write this action like $\sum_{\mathbf{k}}\bar{\Psi}%
_{\mathbf{k}}\mathcal{D}_{\mathbf{k}}\Psi _{\mathbf{k}}$ with the matrix
operator $\mathcal{D}_{\mathbf{k}}$ given by the following $4\times 4$
matrix, 
\begin{equation}
\mathcal{D}_{\mathbf{k}}=\frac{-i}{a}\dsum\limits_{\mu =1}^{4}\gamma ^{\mu }%
\left[ \sin ak_{\mu }-\frac{1}{1+\sqrt{2}}\left( \cos ak_{\mu }-\upsilon
_{\mu }\cos (\dsum\limits_{\nu =1}^{4}ak_{\nu })\right) \right]  \label{AD}
\end{equation}%
with the vector $\upsilon _{\mu }=\left( 1,1,1,1\right) $ as in eq(\ref{VM}%
). In addition to the \emph{four} gamma matrices $\gamma ^{\mu }$, this
operator involves also the particular combination $\sum_{\mu }\gamma ^{\mu
}\upsilon _{\mu }$ required by $SU\left( 5\right) $ symmetry of the
hyperdiamond. By comparing this matrix operator with the generic one (\ref%
{PR}) depending on the complex tensor $\Omega _{\mu }^{l}=R_{\mu
}^{l}+iJ_{\mu }^{l}$ namely, 
\begin{equation}
\mathcal{D}_{\mathbf{k,}\Omega }=\frac{-i}{a}\dsum\limits_{\mu =1}^{4}\gamma
^{\mu }\left[ \dsum\limits_{l=0}^{4}J_{\mu }^{l}\sin
ak_{l}-\dsum\limits_{l=0}^{4}R_{\mu }^{l}\cos ak_{l}\right] ,  \label{DA}
\end{equation}%
we find that (\ref{AD}) is indeed a particular operator of the general (\ref%
{DA}). The corresponding matrices are as follows,%
\begin{equation}
\begin{tabular}{llll}
$\mathrm{R}_{\mu }^{l}=\frac{1}{1+\sqrt{2}}\left( 
\begin{array}{ccccc}
1 & -1 & 0 & 0 & 0 \\ 
1 & 0 & -1 & 0 & 0 \\ 
1 & 0 & 0 & -1 & 0 \\ 
1 & 0 & 0 & 0 & -1%
\end{array}%
\right) $ & $,$ & $\mathrm{J}_{\mu }^{l}=\left( 
\begin{array}{ccccc}
0 & 1 & 0 & 0 & 0 \\ 
0 & 0 & 1 & 0 & 0 \\ 
0 & 0 & 0 & 1 & 0 \\ 
0 & 0 & 0 & 0 & 1%
\end{array}%
\right) $ &  \\ 
&  &  & 
\end{tabular}
\label{JR}
\end{equation}%
Notice that these matrices have no free degrees of freedom, since all
parameters are fixed to defined numbers; they correspond to a particular
point in the moduli space $\frac{U\left( 4,5\right) }{U\left( 4\right)
\times U\left( 5\right) }$ and obey the following remarkable properties 
\begin{equation}
\begin{tabular}{llll}
$\dsum\limits_{l=0}^{4}\mathrm{R}_{\mu }^{l}=0$ & , & $\dsum\limits_{\nu
=1}^{4}\mathrm{J}_{\mu }^{\nu }=\left( 1+\sqrt{2}\right) \mathrm{R}_{\mu
}^{0}$ & $.$%
\end{tabular}
\label{19}
\end{equation}%
These two relations have a physical interpretation; they will be recovered
in next sub-subsection when we study the zero modes of the matrix operator $%
\mathcal{D}_{\mathbf{k,}\Omega }$ for generic $\mathrm{\Omega }_{\mu }^{l}$.
Eqs(\ref{19}) are the conditions that should be obeyed by $\Omega _{\mu
}^{l}=R_{\mu }^{l}+iJ_{\mu }^{l}$ to have two zero modes located at 
\begin{equation}
\begin{tabular}{lll}
$\left( K_{\mu }\right) =\left( \frac{2\pi }{a},\frac{2\pi }{a},\frac{2\pi }{%
a},\frac{2\pi }{a}\right) $ & , & $K_{0}=\frac{2\pi }{a},\qquad \func{mod}%
\frac{2\pi }{a}$ \\ 
$\left( K_{\mu }\right) =\left( \frac{\pi }{4a},\frac{\pi }{4a},\frac{\pi }{%
4a},\frac{\pi }{4a}\right) $ & , & $K_{0}=\frac{-\pi }{a},\qquad \func{mod}%
\frac{2\pi }{a}$%
\end{tabular}%
\end{equation}%
with 
\begin{equation}
\begin{tabular}{lll}
$K_{0}=-\dsum\limits_{\mu =1}^{4}K_{\mu }$ & , & $K_{0}^{\prime
}=-\dsum\limits_{\mu =1}^{4}K_{\mu }^{\prime }$%
\end{tabular}%
\end{equation}%
Notice also that for $K_{\mu }=0$, $\func{mod}\frac{2\pi }{a};$ the operator 
$\mathcal{D}_{\mathbf{K}}$ vanishes exactly due to 
\begin{equation}
\frac{i}{a\left( 1+\sqrt{2}\right) }\sum_{\mu =1}^{4}\gamma ^{\mu }\left(
1-\upsilon _{\mu }\right) =0,
\end{equation}%
and for $K_{\mu }^{\prime }=\frac{\pi }{4a}$, $\func{mod}\frac{2\pi }{a}$,
it also vanishes because of the identity 
\begin{equation}
\frac{-i}{a}\sum_{\mu }\gamma ^{\mu }\left[ \frac{\sqrt{2}}{2}-\frac{1}{1+%
\sqrt{2}}\left( \frac{\sqrt{2}}{2}+\upsilon _{\mu }\right) \right] =0.
\end{equation}%
From these relations, one learns that the operator $\mathcal{D}_{\mathbf{k}}$
has, up to translations in the reciprocal space of the hyperdiamond, two
zero modes located at the points $\mathbf{K}$ and $\mathbf{K}^{\prime }$. To
see that these wave vectors belong indeed to the first Brillouin zone in the
reciprocal space $\mathcal{\tilde{R}}^{4}$, we first use the expansions 
\begin{eqnarray}
&&%
\begin{tabular}{lll}
$\mathbf{K}$ & $=Q_{1}\mathbf{\omega }_{1}+Q_{2}\mathbf{\omega }_{2}+Q_{3}%
\mathbf{\omega }_{3}+Q_{4}\mathbf{\omega }_{4}$ & , \\ 
$\mathbf{K}^{\prime }$ & $=Q_{1}^{\prime }\mathbf{\omega }_{1}+Q_{2}^{\prime
}\mathbf{\omega }_{2}+Q_{3}^{\prime }\mathbf{\omega }_{3}+Q_{4}^{\prime }%
\mathbf{\omega }_{4}$ & ,%
\end{tabular}
\\
&&  \notag
\end{eqnarray}%
with the $Q_{i}$'s and $Q_{i}^{\prime }$'s real numbers; and where the $%
\mathbf{\omega }_{i}$'s\ are the four fundamental weight vectors of SU$%
\left( 5\right) $ that generate $\mathcal{\tilde{R}}^{4}$; they obey amogst
others eqs(\ref{G}-\ref{G1}). Then, we compute the $K_{l}$'s and $%
K_{l}^{\prime }$'s by help of the the relations (\ref{KL}) expressing these
components\ in terms of the $\mathbf{\lambda }_{l}$'s respectively as $%
\mathbf{K}.\mathbf{\lambda }_{l}$ and $\mathbf{K}^{\prime }.\mathbf{\lambda }%
_{l}$. Using eqs(\ref{AL}) giving the $\mathbf{\lambda }_{l}$'s in terms of
the simple roots $\mathbf{\alpha }_{1}$, $\mathbf{\alpha }_{2}$, $\mathbf{%
\alpha }_{3}$, $\mathbf{\alpha }_{4}$; and the duality property $\mathbf{%
\omega }_{i}.\mathbf{\alpha }_{j}=\delta _{ij}$, we can express the
components $k_{l}=\mathbf{k}.\mathbf{\lambda }_{l}$ as $\sum_{i}Q_{i}\left( 
\mathbf{\omega }_{i}.\mathbf{\lambda }_{l}\right) $ or equivalently in
matrix form like,%
\begin{equation}
\left( 
\begin{array}{c}
k_{1} \\ 
k_{2} \\ 
k_{3} \\ 
k_{4}%
\end{array}%
\right) =\left( 
\begin{array}{cccc}
-\frac{1}{5} & +\frac{3}{5} & +\frac{2}{5} & +\frac{1}{5} \\ 
-\frac{1}{5} & -\frac{2}{5} & +\frac{2}{5} & +\frac{1}{5} \\ 
-\frac{1}{5} & -\frac{2}{5} & -\frac{3}{5} & +\frac{1}{5} \\ 
-\frac{1}{5} & -\frac{2}{5} & -\frac{3}{5} & -\frac{4}{5}%
\end{array}%
\right) \left( 
\begin{array}{c}
Q_{1} \\ 
Q_{2} \\ 
Q_{3} \\ 
Q_{4}%
\end{array}%
\right)
\end{equation}%
This leads to%
\begin{equation}
\begin{tabular}{lll}
$Q_{1}=$ & $-2k_{1}-k_{2}-k_{3}-k_{4}$ &  \\ 
$Q_{2}=$ & $k_{1}-k_{2}$ &  \\ 
$Q_{3}=$ & $k_{2}-k_{3}$ &  \\ 
$Q_{4}=$ & $k_{3}-k_{4}$ & 
\end{tabular}%
\end{equation}%
From these relations we learn the two following: (\textbf{i}) the sum $%
k_{1}+k_{2}+k_{3}+k_{4}=-k_{0}$ is precisely $-\frac{1}{5}\left(
4Q_{1}+3Q_{2}+2Q_{3}+Q_{4}\right) $ that follows from the direct use of (\ref%
{AL}). (\textbf{ii}) In the first Brillouin zone given by $\frac{-\pi }{a}%
<k_{\mu }\leq \frac{\pi }{a},$ the two Dirac point $\mathbf{K}$ and $\mathbf{%
K}^{\prime }$ are located at \ 
\begin{equation}
\begin{tabular}{llll}
$\mathbf{K}=0$ & , & $\mathbf{K}^{\prime }=\frac{3\pi }{4a}\mathbf{\omega }%
_{1}=\left( \frac{\pi }{a}-\frac{\pi }{4a}\right) \mathbf{\omega }_{1}$ & .%
\end{tabular}%
\end{equation}%
Notice that in the neighborhood of $\mathbf{K}$ and $\mathbf{K}^{\prime }$,
the operator $\mathcal{D}_{\mathbf{q+K}}$ and $\mathcal{D}_{\mathbf{q+K}%
^{\prime }}$ operators behave, up to the first order in the spacing
parameter $a$, as follows%
\begin{equation}
\begin{tabular}{ll}
$\mathcal{D}_{q+K}=-i\dsum\limits_{\mu =1}^{4}\gamma ^{\mu }q_{\mu }-\frac{ia%
}{2\left( 1+\sqrt{2}\right) }\dsum\limits_{\mu =1}^{4}\gamma ^{\mu }\left[
\left( q_{\mu }\right) ^{2}+\upsilon _{\mu }\left( \mathbf{\upsilon .q}%
\right) ^{2}\right] +O\left( a^{2}\right) $ & , \\ 
$\mathcal{D}_{q+K^{\prime }}=-i\dsum\limits_{\mu =1}^{4}\gamma ^{\mu }q_{\mu
}-\frac{ia}{2\left( 1+\sqrt{2}\right) }\dsum\limits_{\mu =1}^{4}\gamma ^{\mu
}\left[ \frac{\sqrt{2}}{2}\left( k_{\mu }\right) ^{2}-\upsilon _{\mu }\left( 
\mathbf{\upsilon .q}\right) ^{2}\right] +O\left( a^{2}\right) $ & ,%
\end{tabular}
\label{LN}
\end{equation}%
where 
\begin{equation}
\mathbf{\upsilon .q}=\sum_{\nu }\upsilon ^{\nu }q_{\nu
}=q_{1}+q_{2}+q_{3}+q_{4}.
\end{equation}%
These expansions in powers of the spacing parameter of the hyperdiamond
give, at zero order $O\left( a^{0}\right) $, the usual Dirac operator of the
continuum limit. At first order $O\left( a\right) $, we have 5-dimensional
operators that play an important in the study of the Symanzik theory. The
latter has a generic effective action that reads in powers of the parameter $%
a$ as follows,%
\begin{equation}
\mathcal{S}_{eff}=\frac{1}{a^{4}}\sum_{n}a^{n}\sum_{j}c_{n}^{\left( j\right)
}\mathcal{O}_{n}^{\left( j\right) }.
\end{equation}%
Here, the $\mathcal{O}_{n}^{\left( j\right) }$'s are dimension $n$ operators
with relevant $\mathcal{O}_{3}^{\left( j\right) }$ and marginal $\mathcal{O}%
_{4}^{\left( j\right) }$ ones; some of these operators may emerge in the
renormalization of the quantum theory. The last terms of (\ref{LN}) are
linear in $a$ and so lead to $\mathcal{O}_{5}^{\left( j\right) }$ operators;
for explicit use of these operators; see \textrm{\cite{12}}.

\subsection{Generic solutions}

In this subsection we want to further explore the role of the complex tensor 
$\Omega _{\mu }^{l}$ in the engineering of zero modes of the matrix operator 
$\mathcal{D}_{\mathbf{k,}\Omega }$ of eq(\ref{PR}). We start from eq(\ref{DA}%
); then look for the interesting class of those $\mathcal{D}_{\mathbf{k,}%
\Omega }$ operators having:

\begin{itemize}
\item two zero modes located at two different points $\mathbf{K}$ and $%
\mathbf{K}^{\prime }$ of the reciprocal space. These modes, which should be
non degenerate, require the two set of conditions:%
\begin{equation}
\begin{tabular}{llll}
$\mathcal{D}_{\mathbf{K,}\Omega }=0$ & and & $\mathcal{D}_{\mathbf{K}%
^{\prime }\mathbf{,}\Omega }=0$ & ,%
\end{tabular}
\label{KD}
\end{equation}

\item the appropriate continuum limit in neighborhood of $K$ and $K^{\prime
} $ given by the Dirac operator. From eq(\ref{DA}), this property requires
amongst others that the imaginary part of $\Omega _{\mu }^{l}$ should be as $%
J_{\mu }^{l}\neq A_{\mu }\otimes B^{l}$,

\item the $SU\left( 5\right) $ symmetry of the hyperdiamond to take
advantage of known results on \emph{2D}\ graphene.
\end{itemize}

\ \newline
To avoid lenghty technical details on the engineering of the zero modes, we
will come directly to the key points of our method of construction. The
basic ideas are illustrated through explicit examples given in next
sub-subsection; and general results will be given later.

\subsubsection{Method of engineering zeros}

First, notice that the values of the wave vector variable $\mathbf{k}$
solving $\mathcal{D}_{\mathbf{k,}\Omega }=0$ depend on the components of $%
\Omega _{\mu }^{l}$; this feature means that this zero mode equations can be
also interpreted as constraint eqs relating wave vectors $k_{\mu }$ in the
reciprocal lattice $\mathcal{\tilde{R}}^{4}$ to points in the real \emph{40}%
- dimensional moduli space $\mathfrak{M}_{40}=\mathfrak{R}_{20}\times 
\mathfrak{R}_{20}$ given by,%
\begin{equation}
\mathfrak{M}_{40}=\frac{SO\left( 4,5\right) }{SO\left( 4\right) \times
SO\left( 5\right) }\times \frac{SO\left( 4,5\right) }{SO\left( 4\right)
\times SO\left( 5\right) }.
\end{equation}%
The relations between the wave vectors $\mathbf{k}$ and the moduli $\Omega
_{\mu }^{l}$ mean as well that we have the two following equivalent
statements:

\begin{itemize}
\item Given some tensor $\Omega _{\mu }^{l}$, corresponding to a point in
the moduli space $\mathfrak{M}_{40}$, the solving of $\mathcal{D}_{\mathbf{k,%
}\Omega }=0$ lead to wave vectors $\left\{ K_{\mu }^{\left( 1\right)
},K_{\mu }^{\left( 2\right) },...\right\} $ in the reciprocal space $%
\mathcal{\tilde{R}}^{4}$. This approach has been used in the above
subsection with $\Omega _{\mu }^{l}$ given by eqs(\ref{JR}).

\item Inversely, given a wave vector $K_{\mu }$, corresponding to point in $%
\mathcal{\tilde{R}}^{4}$, the matrix equation $\mathcal{D}_{\mathbf{K,}%
\Omega }=0$ allow to determine the set of points in $\mathfrak{M}_{40}$. In
other words, $\mathcal{D}_{\mathbf{K,}\Omega }$ should be thought of as
constraint eqs on the \emph{40} real moduli of the complex tensor $\Omega
_{\mu }^{l}=R_{\mu }^{l}+iJ_{\mu }^{l}$. \newline
Below, we develop this second approach.
\end{itemize}

\emph{Zero mode's constraint eqs and their linearization}\newline
To do so, notice that $\mathcal{D}_{\mathbf{K,}\Omega }=0$ is a $4\times 4$
matrix operator equation that describe \emph{a priori} a system of \emph{16}
eqs; but because of the properties of the Dirac matrices, only the \emph{4}
of these relations are relevant; they are given by,%
\begin{equation}
Tr\left( \gamma _{\mu }\mathcal{D}_{\mathbf{K,}\Omega }\right) =0,\qquad \mu
=1,2,3,4.
\end{equation}%
Notice also that in the interesting case of two zero modes of $\mathcal{D}_{%
\mathbf{k,}\Omega }$, we need to fix two points in the reciprocal space; say 
$\mathbf{K}=\left( K_{1},K_{2},K_{3},K_{4}\right) $ and $\mathbf{K}^{\prime
}=\left( K_{1}^{\prime },K_{2}^{\prime },K_{3}^{\prime },K_{4}^{\prime
}\right) $ satisying the conditions,%
\begin{equation}
Tr\left( \gamma _{\mu }\mathcal{D}_{\mathbf{K,}\Omega }\right) =0,\qquad
Tr\left( \gamma _{\mu }\mathcal{D}_{\mathbf{K}^{\prime }\mathbf{,}\Omega
}\right) =0.
\end{equation}%
These conditions constitute \emph{8} constraints on the free parameters of
the complex tensor $\Omega _{\mu }^{l}$; their solutions require fixing 8
parameters of $\Omega _{\mu }^{l}$ in terms of $\left(
K_{1},K_{2},K_{3},K_{4}\right) $ and $\left( K_{1}^{\prime },K_{2}^{\prime
},K_{3}^{\prime },K_{4}^{\prime }\right) $; and therefore lead to a
reduction of the moduli space $\mathfrak{M}_{40}$ down to some subspaces $%
\mathfrak{M}_{32}$,%
\begin{equation}
\mathfrak{M}_{40}\rightarrow \mathfrak{M}_{32}.
\end{equation}%
The mathematical structure of $\mathfrak{M}_{32}$ depends obviously on the
nature of the constraints on $\Omega _{\mu }^{l}$. To have an idea on the
space $\mathfrak{M}_{32}$, we study below some explicit examples where $%
\mathfrak{M}_{32}$ is, for instance, given by the complex coset group
manifold $\frac{U\left( 4,4\right) }{U\left( 4\right) \times U\left(
4\right) }$ or by the following real one $\frac{SO\left( 4,3\right) }{%
SO\left( 4\right) \times SO\left( 3\right) }\times \frac{SO\left( 4,5\right) 
}{SO\left( 4\right) \times SO\left( 5\right) }$. Recal by the way that $\dim
U\left( N,M\right) =$ $\left( N+M\right) ^{2}$ and $\dim SO\left( N,M\right)
=$ $\frac{1}{2}\left( N+M\right) \left( M+N-1\right) $.\newline
Substituting the decomposition $\Omega _{\mu }^{l}=R_{\mu }^{l}+iJ_{\mu
}^{l} $ back into (\ref{AD}), we get the following constraint relation for
zero modes,%
\begin{equation}
\mathcal{D}_{\mathbf{k,}\Omega }=\frac{-i}{a}\dsum\limits_{\mu =1}^{4}\gamma
^{\mu }\left( \dsum\limits_{l=0}^{4}J_{\mu }^{l}\sin
ak_{l}-\dsum\limits_{l=0}^{4}R_{\mu }^{l}\cos ak_{l}\right) =0,  \label{DD}
\end{equation}%
where the $ak_{l}=\Phi _{l}$ are the phases of the wave propagation, along
the $\mathbf{\lambda }_{l}$\ direction in the reciprocal space, satisfying
the SU$\left( 5\right) $ symmetry condition,%
\begin{equation}
\begin{tabular}{llll}
$\dsum\limits_{l=0}^{4}k_{l}=0,$ $\ \func{mod}\frac{2\pi }{a}$\  & $,$ & $%
\dsum\limits_{l=0}^{4}\Phi _{l}=0,$ \ $\func{mod}2\pi $ & .%
\end{tabular}
\label{CON}
\end{equation}%
Using the diagonal vector $\mathbf{\upsilon =}\left( 1,1,1,1\right) $, these
constraints allows to express $k_{0}$ in terms of the others; and similarly
for $\Phi _{0}$. We have: 
\begin{equation}
\begin{tabular}{llll}
$k_{0}=-\dsum\limits_{\mu =1}^{4}k_{\mu }\equiv -\mathbf{k.\upsilon }$ & $,$
& $\Phi _{0}=-\dsum\limits_{\mu =1}^{4}\Phi _{\mu }\equiv -\mathbf{\Phi
.\upsilon }$ & .%
\end{tabular}%
\end{equation}%
Notice that eqs(\ref{DD}) is a system 4 equations with 4 variables $\Phi
_{1},$ $\Phi _{2},$ $\Phi _{3},$ $\Phi _{4}$ and \emph{40} real moduli given
by the composantes of $R_{\mu }^{l}$ and $J_{\mu }^{l}$. Moreover because of
their dependence on the functions $\cos \Phi _{\mu }$ and $\sin \Phi _{\mu }$%
; and also because of the condition (\ref{CON}), these numerical eqs are
highly non linear and therefore difficult to solve. To reduce this non
linearity, it is interesting to work with the following new variables, 
\begin{equation}
\begin{tabular}{llll}
$C_{\mu }=\cos \Phi _{\mu }$ & , & $S_{\mu }=\sin \Phi _{\mu }$ & , \\ 
$\vartheta _{\mu }=R_{\mu }^{0}C_{0}$ & , & $\varphi _{\mu }=J_{\mu
}^{0}S_{0}$ & ,%
\end{tabular}
\label{SC}
\end{equation}%
this allow to gain some simplicity into the zero mode constraint eqs.
Focusing on the first Brillouin zone in the reciprocal space $\frac{-\pi }{a}%
\leq k_{\mu }\leq \frac{\pi }{a}$; and using the above change of variables
we can put (\ref{DD}) into the form%
\begin{equation}
\begin{tabular}{ll}
$\dsum\limits_{\nu =1}^{4}\left( R_{\mu }^{\nu }C_{\nu }-J_{\mu }^{\nu
}S_{\nu }\right) =J_{\mu }^{0}S_{0}-R_{\mu }^{0}C_{0}$ & ,%
\end{tabular}
\label{3D}
\end{equation}%
but still with the non linear terms%
\begin{equation}
\begin{tabular}{lll}
$C_{0}=$ & $%
C_{1}C_{2}C_{3}C_{4}-C_{1}C_{2}S_{3}S_{4}-C_{1}S_{2}C_{3}S_{4}-C_{1}S_{2}S_{3}C_{4} 
$ &  \\ 
& $%
-S_{1}C_{2}C_{3}S_{4}-S_{1}C_{2}S_{3}C_{4}-S_{1}S_{2}C_{3}C_{4}+S_{1}S_{2}S_{3}S_{4} 
$ & 
\end{tabular}%
\end{equation}%
and similarly for others coming from the expansion of $S_{0}$. We will see
below how these non linearities can be overcome.

\emph{Solving eqs(\ref{KD})}\newline
The solutions we are interested in here are those satisfying the \emph{3}
properties given in the begining of this subsection namely: (i) two zeros
modes at $\mathbf{K}$ and $\mathbf{K}^{\prime }$ of the reciprocal space,
(ii) a Dirac operator in the continuum limit of the neighborhood of $\mathbf{%
K}$ and $\mathbf{K}^{\prime }$, and (iii) SU$\left( 5\right) $ symmetry.%
\newline
Our method to solve eqs(\ref{KD}) involves \emph{4} steps as follows:\newline
(\textbf{1}) Start from the generic zero mode constraint equations $\mathcal{%
D}_{\mathbf{k,}\Omega }=0$ given by (\ref{3D}), \ \ \newline
(\textbf{2}) Take two arbitrary wave vectors $\mathbf{K}$ and $\mathbf{K}%
^{\prime }$ in the first Brillouin zone of the reciprocal space, which by
help of eqs(\ref{SC}), can be related to $C_{l}$ and $S_{l}$ like, 
\begin{equation}
\begin{tabular}{llll}
$e^{iaK.\lambda _{l}}=C_{l}+iS_{l}$ & , & $e^{iaK^{\prime }.\lambda
_{l}}=C_{l}^{\prime }+iS_{l}^{\prime }$ & ,%
\end{tabular}%
\end{equation}%
leading in turn to the following system of \emph{4+4} relations defining the
two zeros%
\begin{equation}
\begin{tabular}{ll}
$\dsum\limits_{\nu =1}^{4}\left( R_{\mu }^{\nu }C_{\nu }-J_{\mu }^{\nu
}S_{\nu }\right) =J_{\mu }^{0}S_{0}-R_{\mu }^{0}C_{0}$ & , \\ 
$\dsum\limits_{\nu =1}^{4}\left( R_{\mu }^{\nu }C_{\nu }^{\prime }-J_{\mu
}^{\nu }S_{\nu }^{\prime }\right) =J_{\mu }^{0}S_{0}^{\prime }-R_{\mu
}^{0}C_{0}^{\prime }$ & .%
\end{tabular}
\label{PC}
\end{equation}%
with variables $C_{\nu }$ and $C_{\nu }^{\prime }$ (since $S_{\nu }=\pm 
\sqrt{1-C_{\nu }^{2}}$, $S_{\nu }^{\prime }=\pm \sqrt{1-C_{\nu }^{\prime 2}}$%
) and the free moduli $R_{\mu }^{\nu }$, $R_{\mu }^{0}$ as well as $J_{\mu
}^{\nu }$, $J_{\mu }^{0}$. \newline
(\textbf{3}) To surround the non linearities, we make a particular choice of 
$K_{\mu }$ and $K_{\mu }^{\prime }$ in the reciprocal space and interpret (%
\ref{PC}) as constraint equations on the moduli. For example, one may take
the wave vectors $K_{\mu }$ and $K_{\mu }^{\prime }$ as follows,%
\begin{equation}
\begin{tabular}{llll}
$K_{\mu }=\left( 0,0,0,0\right) $ & , & $K_{\mu }^{\prime }=\left( \frac{\pi 
}{4a},\frac{\pi }{4a},\frac{\pi }{4a},\frac{\pi }{4a}\right) $ & .%
\end{tabular}
\label{ZMO}
\end{equation}%
These vectors are precisely the one that have been obtained in the previous
subsection; but one can also consider other choices; some of them are
described as remarks which will be given at the end of this construction.
Using the above particular choice, we have the following:%
\begin{equation}
\begin{tabular}{lllll}
$K_{0}=0$ & , & $\left( C_{0},S_{0}\right) =\left( +1,0\right) $ & , & $%
\left( C_{\mu },S_{\mu }\right) =\left( 1,0\right) $ \\ 
$K_{0}^{\prime }=-\frac{\pi }{a}$ & , & $\left( C_{0}^{\prime
},S_{0}^{\prime }\right) =\left( -1,0\right) $ & , & $\left( C_{\mu
}^{\prime },S_{\mu }^{\prime }\right) =\left( \frac{\sqrt{2}}{2},\frac{\sqrt{%
2}}{2}\right) $%
\end{tabular}
\label{ZM}
\end{equation}%
(\textbf{4}) Put these values back into eqs(\ref{PC}), we obtain the
following constraint relations on the tensors $R_{\mu }^{l}$ and $J_{\mu
}^{l},$ 
\begin{equation}
\begin{tabular}{ll}
$R_{\mu }^{0}+\dsum\limits_{\nu =1}^{4}R_{\mu }^{\nu }=0$, & $%
\dsum\limits_{\nu =1}^{4}J_{\mu }^{\nu }=-\left( 1+\sqrt{2}\right) R_{\mu
}^{0},$%
\end{tabular}
\label{E3}
\end{equation}%
which should be compared with (\ref{19}). The first set of constraint eqs
reduce the number of degrees of freedom of $R_{\mu }^{l}$ from \emph{20}
down to \emph{16}. The second set of constraints do the same thing; but with
the tensor $J_{\mu }^{l}$. So the moduli space $\mathfrak{M}_{40}=\mathfrak{R%
}_{20}\times \mathfrak{R}_{20}$ gets reduced down to the subspace%
\begin{equation}
\mathfrak{M}_{32}=\frac{SO\left( 4,4\right) }{SO\left( 4\right) \times
SO\left( 4\right) }\times \frac{SO\left( 4,4\right) }{SO\left( 4\right)
\times SO\left( 4\right) }.
\end{equation}%
Therefore, the two zeros (\ref{ZMO}) live on $\mathfrak{M}_{32}$; those
missing \emph{8} moduli living on the cosed manifold $\mathfrak{M}_{40}/%
\mathfrak{M}_{32}$ has been freezed by eqs(\ref{ZMO}).\newline
With this method, one can engineer various 4D lattice models living on
hyperdiamond. In this regards, let us consider rapidly other examples; and
turn after to give a general result through three claims.

\emph{A remark and other examples}\newline
The remark concerns the general form of the operator $\mathcal{D}_{\mathbf{k,%
}\Omega }$ that has two zero modes located at the points $K_{\mu }=\left(
0,0,0,0\right) $ and $K_{\mu }^{\prime }=\left( \frac{\pi }{4a},\frac{\pi }{%
4a},\frac{\pi }{4a},\frac{\pi }{4a}\right) $. It is given by:%
\begin{equation}
\begin{tabular}{lll}
$\mathcal{D}_{\mathbf{k,}\Omega }=$ & $\frac{-i}{a}\dsum\limits_{\mu
=1}^{4}\gamma ^{\mu }\left( \dsum\limits_{\nu =1}^{4}J_{\mu }^{\nu }\sin
ak_{\nu }-\dsum\limits_{\nu =1}^{4}R_{\mu }^{\nu }\cos ak_{\nu }\right) $ & 
\\ 
& $\frac{-i}{a}\dsum\limits_{\mu =1}^{4}\gamma ^{\mu }\left[ -J_{\mu
}^{0}\sin a\mathbf{k.\upsilon +}\left( \dsum\limits_{\nu =1}^{4}R_{\mu
}^{\nu }\right) \cos a\mathbf{k.\upsilon }\right] ,$ & 
\end{tabular}%
\end{equation}%
with the condition 
\begin{equation}
\dsum\limits_{\nu =1}^{4}J_{\mu }^{\nu }=\left( 1+\sqrt{2}\right)
\dsum\limits_{\nu =1}^{4}R_{\mu }^{\nu }.
\end{equation}%
Notice that in the neighborhood of $K_{\mu }$, the expansion of $\mathcal{D}%
_{\mathbf{K+q,}\Omega }$ reads, up to second order in the power of the
lattice spacing parameter $a$, like: 
\begin{equation}
\begin{tabular}{lll}
$\mathcal{D}_{\mathbf{q,}\Omega }=$ & $-i\dsum\limits_{\mu =1}^{4}\gamma
^{\mu }\left( \dsum\limits_{\nu =1}^{4}\left[ J_{\mu }^{\nu }-\upsilon ^{\nu
}J_{\mu }^{0}\right] q_{\nu }\right) $ &  \\ 
& $\frac{-ai}{2}\dsum\limits_{\mu =1}^{4}\gamma ^{\mu }\left[
\dsum\limits_{\nu =1}^{4}\left( q_{\nu }\right) ^{2}R_{\mu }^{\nu }\mathbf{+}%
\left( \mathbf{q.\upsilon }\right) ^{2}R_{\mu }^{0}\right] +O\left(
a^{2}\right) ,$ & 
\end{tabular}%
\end{equation}%
where we have used $\sum_{\nu =1}^{4}R_{\mu }^{\nu }=-R_{\mu }^{0}$. To
interpret $q_{\mu }$ as the real wave vector of the Dirac theory, we have to
require moreover 
\begin{equation}
J_{\mu }^{\nu }-\upsilon ^{\nu }J_{\mu }^{0}\sim \delta _{\mu }^{\nu }.
\end{equation}%
The other examples concern different kinds of two zero modes other than the
ones given by (\ref{ZMO}). For instance, in the case where the two zero
modes are chosen like in \emph{BC} model (\ref{BCM}) namely 
\begin{equation}
\begin{tabular}{llll}
$\Phi _{1}=\Phi _{2}=\Phi _{3}=\Phi _{4}=0$ & , & $\Phi _{0}=0$ & , \\ 
$\Phi _{1}^{\prime }=\Phi _{2}^{\prime }=\Phi _{3}^{\prime }=\Phi
_{4}^{\prime }=\frac{\pi }{2}$ & , & $\Phi _{0}^{\prime }=-2\pi $ & ,%
\end{tabular}%
\end{equation}%
the analogue of eqs(\ref{ZM}) read like,%
\begin{equation}
\begin{tabular}{lllll}
$K_{0}=0$ & , & $\left( C_{0},S_{0}\right) =\left( 1,0\right) $ & , & $%
\left( C_{\mu },S_{\mu }\right) =\left( 1,0\right) $ \\ 
$K_{0}^{\prime }=-\frac{2\pi }{a}$ & , & $\left( C_{0}^{\prime
},S_{0}^{\prime }\right) =\left( 1,0\right) $ & , & $\left( C_{\mu }^{\prime
},S_{\mu }^{\prime }\right) =\left( 0,1\right) $%
\end{tabular}%
\end{equation}%
Putting these values back into (\ref{PC}), we get the following constraint
relations,%
\begin{equation}
\begin{tabular}{llll}
$\dsum\limits_{\nu =1}^{4}R_{\mu }^{\nu }=-R_{\mu }^{0}$ & , & $%
\dsum\limits_{\nu =1}^{4}J_{\mu }^{\nu }=R_{\mu }^{0}$ & ,%
\end{tabular}
\label{BCB}
\end{equation}%
which can be also put into the form 
\begin{equation}
\sum_{\nu =1}^{4}\left( R_{\mu }^{\nu }+iJ_{\mu }^{\nu }\right) =\left(
-1+i\right) R_{\mu }^{0}.
\end{equation}%
These condition are manifestly satisfied by the Bori\c{c}i-Creutz matrix
given by eq(\ref{BC}) and the moduli space of the tensor $\Omega _{\mu }^{l}$
is given by $\frac{U\left( 4,4\right) }{U\left( 4\right) \times U\left(
4\right) }$. Similarly, in the case the two zero modes are chosen like 
\begin{equation}
\begin{tabular}{llll}
$\Phi _{1}=\Phi _{2}=\Phi _{3}=0,$ & $\Phi _{4}=0$, & $\Phi _{0}=0$ & , \\ 
$\Phi _{1}^{\prime }=\Phi _{2}^{\prime }=\Phi _{3}^{\prime }=0,$ & $\Phi
_{4}^{\prime }=\pi $, & $\Phi _{0}^{\prime }=-\pi $ & ,%
\end{tabular}%
\end{equation}%
as in the \emph{KW} fermions, the analogue of eqs(\ref{ZM}) read like,%
\begin{equation}
\begin{tabular}{lllll}
$K_{0}=0$ & , & $\left( C_{0},S_{0}\right) =\left( +1,0\right) $ & , & $%
\left( C_{4},S_{4}\right) =\left( +1,0\right) $ \\ 
$K_{0}^{\prime }=0$ & , & $\left( C_{0}^{\prime },S_{0}^{\prime }\right)
=\left( -1,0\right) $ & , & $\left( C_{4}^{\prime },S_{4}^{\prime }\right)
=\left( -1,0\right) $%
\end{tabular}%
,
\end{equation}%
and for $\mu =1,2,3$, we have%
\begin{equation}
\begin{tabular}{ll}
$\left( C_{\mu },S_{\mu }\right) =\left( 1,0\right) $ & , \\ 
$\left( C_{\mu }^{\prime },S_{\mu }^{\prime }\right) =\left( 1,0\right) $ & .%
\end{tabular}%
\end{equation}%
The constraint relations (\ref{PC}) become,%
\begin{equation}
\begin{tabular}{llll}
$R_{\mu }^{0}+\dsum\limits_{\nu =1}^{4}R_{\mu }^{\nu }=0$ & , & $R_{\mu
}^{1}+R_{\mu }^{2}+R_{\mu }^{3}=R_{\mu }^{0}+R_{\mu }^{4}$ & ,%
\end{tabular}
\label{KWK}
\end{equation}%
exactly as for the \emph{KW} matrix given by eq(\ref{KW}); see also next
section. Notice that, in this example, the \emph{4+4} constraint relations
are all of them imposed on $R_{\mu }^{l}$, the real part of $\Omega _{\mu
}^{l}$; so the moduli space associated with these classes of solutions is
given by 
\begin{equation}
\mathfrak{M}_{32}^{\prime }=\frac{SO\left( 4,3\right) }{SO\left( 4\right)
\times SO\left( 3\right) }\times \frac{SO\left( 4,5\right) }{SO\left(
4\right) \times SO\left( 5\right) }
\end{equation}%
Below, we give three claims summarizing the basic results following from
this construction.

\subsubsection{three claims}

\begin{claim}
: Moduli space of $\mathcal{D}_{\mathbf{k,}\Omega }$\newline
The moduli space of the operator $\mathcal{D}_{\mathbf{k,}\Omega }\sim \frac{%
1}{a}\sum_{\mu }\gamma ^{\mu }\left( \sum_{l}\Omega _{\mu }^{l}e^{ia\mathbf{k%
}.\mathbf{\lambda }_{l}}+hc\right) $ on 4D hyperdiamond is given by the
complexification of the real 20-dimensional space $\mathfrak{R}_{20}=\frac{%
SO\left( 4,5\right) }{SO\left( 4\right) \times SO\left( 5\right) }$. Each
copy of the two $\mathfrak{R}_{20}$'s is associated with the real $R_{\mu
}^{l}$ and imaginary $J_{\mu }^{l}$ parts of the complex tensor $\Omega
_{\mu }^{l}$. The combination of these spaces leads to the following complex
20- dimensional coset group manifold 
\begin{equation}
\mathfrak{C}_{20}=\frac{U\left( 4,5\right) }{U\left( 4\right) \times U\left(
5\right) }\sim \mathfrak{M}_{40}
\end{equation}
\end{claim}

\begin{claim}
: Engineering zero modes\newline
Each zero mode of $\mathcal{D}_{\mathbf{k,}\Omega },$ located at some wave
vector point $\mathbf{K=}\left( K_{1},K_{2},K_{3},K_{4}\right) $ in the
reciprocal space of the 4D hyperdiamond, requires four real constraint
relations given by the vanishing conditions $Tr\left[ \gamma _{\mu }\mathcal{%
D}_{\mathbf{k,}\Omega }\right] =0$. These constraints reduce the moduli
space $\mathfrak{M}_{40}$ down to some submanifolds $\mathfrak{M}_{36}$.
Similarly, two zero modes, located at two points $\mathbf{K}$ and $\mathbf{K}%
^{\prime }$ in the reciprocal space, require 4 complex ($4+4$ real)
conditions given by 
\begin{equation}
\begin{tabular}{llll}
$Tr\left[ \gamma _{\mu }\mathcal{D}_{K,\Omega }\right] =0$ & , & $Tr\left[
\gamma _{\mu }\mathcal{D}_{K^{\prime },\Omega }\right] =0$ & .%
\end{tabular}%
\end{equation}%
These constraint relations reduce the moduli space $\mathfrak{M}_{40}$ down
to real 32-dimensional subspaces $\mathfrak{M}_{32}$. In the case of \emph{BC%
} fermions, the $\mathfrak{M}_{32}$ manifold is given by 
\begin{equation}
\mathfrak{C}_{16}=\frac{U\left( 4,4\right) }{U\left( 4\right) \times U\left(
4\right) },
\end{equation}%
and in \emph{KW} fermions; it reads as $\frac{SO\left( 4,3\right) }{SO\left(
4\right) \times SO\left( 3\right) }\times \frac{SO\left( 4,5\right) }{%
SO\left( 4\right) \times SO\left( 5\right) }.$ For the generic case of $%
\left( n+m\right) $ zero modes of $\mathcal{D}_{K,\Omega }$, one should have 
$4\left( n+m\right) $ conditions; and the corresponding moduli spaces have
the form $\mathfrak{M}_{40-4n-4m}$. Typical examples of these submanifolds
are given by $\mathfrak{R}_{20-4n}\times \mathfrak{R}_{20-4m}$ with $%
\mathfrak{R}_{20-4s}=\frac{SO\left( 4,5-s\right) }{SO\left( 4\right) \times
SO\left( 5-s\right) }$.
\end{claim}

\begin{claim}
: Continuum limit of $\mathcal{D}_{q+K\mathbf{,}\Omega }$\newline
In the neighborhood of each zero mode $K$ with $Tr\left[ \gamma _{\mu }%
\mathcal{D}_{K,\Omega }\right] =0$, the operator $\mathcal{D}_{q+K,\Omega }$
behaves, at first order in the lattice spacing parameter a, like%
\begin{equation}
\begin{tabular}{lll}
$\mathcal{D}_{q+K,\Omega }\sim $ & $-\dsum\limits_{\mu =1}^{4}\gamma ^{\mu
}p_{\mu }$ &  \\ 
& $-a\dsum\limits_{\mu =1}^{4}\gamma ^{\mu }\left[ R_{\mu }^{l}\cos
aK_{l}-J_{\mu }^{l}\sin aK_{l}\right] \left( q_{l}\right) ^{2}+O\left(
a^{2}\right) $ & 
\end{tabular}%
\end{equation}%
where the wave vector $p_{\mu }$ is related to $q_{l}$ as follows 
\begin{equation}
p_{\mu }=\sum_{\nu =1}^{4}\left( R_{\mu }^{\nu }S_{\nu }+J_{\mu }^{\nu
}S_{\nu }\right) q_{\nu }+\left( R_{\mu }^{0}S_{0}+J_{\mu }^{0}C_{0}\right)
q_{0},
\end{equation}%
with $q_{l}=\mathbf{q}.\mathbf{\lambda }_{l}$, $K_{l}=\mathbf{K}.\mathbf{%
\lambda }_{l}$ and $C_{l}=\cos aK_{l}$, $S_{l}=\sin aK_{l}$. To interpret
the $q_{\mu }$'s as the real wave vector components of the continuum limit,
we have to impose moreover $\left( R_{\mu }^{\nu }S_{\nu }+J_{\mu }^{\nu
}C_{l}\right) \sim \delta _{\mu }^{\nu }$ and $\left( R_{\mu
}^{0}S_{0}+J_{\mu }^{0}C_{0}\right) =0$.
\end{claim}

\section{Re-deriving \emph{BC} and \emph{KW} fermions}

In this section, we use $SU\left( 5\right) $ symmetry of the hyperdiamond to
re-derive the \emph{BC} fermions and the \emph{KW} ones from our general
action. We first consider \emph{BC} model and then the \emph{KW} fermions

\subsection{BC fermions}

A simple way to re-derive the \emph{BC} fermions from our proposal is to
work in the reciprocal space and compare the Dirac operators following from
the models (\ref{BBC}) and (\ref{PR}). In the case of the \emph{BC}
fermions, the Dirac operator $\mathcal{D}_{BC}$ reads, up to a scale factor
and upon dropping the mass term, as follows%
\begin{equation}
\mathcal{D}_{BC}\text{ }\sim \text{ }\frac{i}{a}\dsum\limits_{\mu
=1}^{4}\gamma ^{\mu }\sin ap_{\mu }-\frac{2i}{a}\Gamma -\frac{i}{a}%
\dsum\limits_{\mu =1}^{4}\left( \gamma ^{\mu }-\Gamma \right) \cos ap_{\mu },
\label{A1}
\end{equation}%
with 
\begin{equation}
\begin{tabular}{llll}
$\Gamma $ & $=$ & $\frac{1}{2}\left( \gamma ^{1}+\gamma ^{2}+\gamma
^{3}+\gamma ^{4}\right) $ & ,%
\end{tabular}%
\end{equation}%
which we rewrite as 
\begin{equation}
\begin{tabular}{llll}
$\Gamma =\frac{1}{2}\dsum\limits_{\mu =1}^{4}\gamma ^{\mu }\upsilon _{\mu }$
& , & $\upsilon _{\mu }=\left( 
\begin{array}{c}
1 \\ 
1 \\ 
1 \\ 
1%
\end{array}%
\right) $ & .%
\end{tabular}
\label{A3}
\end{equation}%
In our proposal (\ref{PR}), the Dirac operator reads as%
\begin{equation}
\begin{tabular}{ll}
$\mathcal{D}\text{ }\sim \text{ }$ & $\dsum\limits_{\mu =1}^{4}\gamma ^{\mu }%
\left[ \Omega _{\mu }^{0}-\bar{\Omega}_{\mu }^{0}\right] \sin
ap_{0}+\dsum\limits_{\mu ,\nu =1}^{4}\gamma ^{\mu }\left[ \Omega _{\mu
}^{\nu }-\bar{\Omega}_{\mu }^{\nu }\right] \sin ap_{\nu }$ \\ 
& $-i\dsum\limits_{\mu =1}^{4}\gamma ^{\mu }\left[ \Omega _{\mu }^{0}+\bar{%
\Omega}_{\mu }^{0}\right] \cos ap_{0}-i\dsum\limits_{\mu ,\nu =1}^{4}\gamma
^{\mu }\left[ \Omega _{\mu }^{\nu }+\bar{\Omega}_{\mu }^{\nu }\right] \cos
ap_{\nu }.$ \\ 
& 
\end{tabular}
\label{A4}
\end{equation}%
By equating (\ref{A1}) and (\ref{A4}), one gets the tensor $\Omega _{\mu
}^{l}$ that define the \emph{BC} model. Taking the components of the tensor $%
\Omega _{\mu }^{l}$ as follows%
\begin{equation}
\Omega _{\mu }^{l}\neq \bar{\Omega}_{\mu }^{l}
\end{equation}%
with%
\begin{equation}
\begin{tabular}{llll}
$\Omega _{\mu }^{0}=\frac{1}{2}\xi _{\mu }$ & , & $\Omega _{\mu }^{\nu }=%
\frac{1}{2}\left( 1+i\right) \delta _{\mu }^{\nu }-\frac{1}{4}\Sigma _{\mu
}^{\nu }$ & ,%
\end{tabular}%
\end{equation}%
and putting back into (\ref{A4}), we first get%
\begin{equation}
\begin{tabular}{ll}
$\mathcal{D}=$ & $\frac{1}{2a}\dsum\limits_{\mu =1}^{4}\gamma ^{\mu }\left[
\xi _{\mu }-\bar{\xi}_{\mu }\right] \sin ap_{0}+\frac{i}{2a}%
\dsum\limits_{\mu =1}^{4}\gamma ^{\mu }\sin ap_{\mu }$ \\ 
& $-\frac{i}{2a}\dsum\limits_{\mu =1}^{4}\gamma ^{\mu }\left[ \xi _{\mu }+%
\bar{\xi}_{\mu }\right] \cos ap_{0}-\frac{i}{2a}\dsum\limits_{\mu ,\nu
=1}^{4}\gamma ^{\mu }\left[ 2\delta _{\mu }^{\nu }-\Sigma _{\mu }^{\nu }%
\right] \cos ap_{\nu }.$%
\end{tabular}
\label{A5}
\end{equation}%
Then equating with (\ref{A1}), we obtain:%
\begin{equation}
\begin{tabular}{llllll}
$\xi _{\mu }-\bar{\xi}_{\mu }$ & $=0$ & , & or & $\sin ap_{0}=0$ & ,%
\end{tabular}%
\end{equation}%
and%
\begin{equation}
\begin{tabular}{lll}
$\Omega _{\mu }^{\nu }-\bar{\Omega}_{\mu }^{\nu }$ & $=i\delta _{\mu }^{\nu
} $ & , \\ 
$\Omega _{\mu }^{\nu }+\bar{\Omega}_{\mu }^{\nu }$ & $=\delta _{\mu }^{\nu }-%
\frac{1}{2}\Sigma _{\mu }^{\nu }$ & ,%
\end{tabular}%
\end{equation}%
where $\xi _{\mu }$ and $\Sigma _{\mu }^{\nu }$ are still to determine.
Solving these quantities as follows,%
\begin{equation}
\begin{tabular}{llll}
$\xi _{\mu }=\upsilon _{\mu }$ & , & $\Sigma _{\mu }^{\nu }=\upsilon _{\mu
}\otimes \upsilon ^{\nu }$ & ,%
\end{tabular}%
\end{equation}%
where $\upsilon _{\mu }$ is as in (\ref{A3}) and where 
\begin{equation}
\Sigma _{\mu }^{\nu }=\left( 
\begin{array}{cccc}
1 & 1 & 1 & 1 \\ 
1 & 1 & 1 & 1 \\ 
1 & 1 & 1 & 1 \\ 
1 & 1 & 1 & 1%
\end{array}%
\right)
\end{equation}%
then putting back into (\ref{A5}), we end with the following factorized form%
\begin{equation}
\begin{tabular}{ll}
$\mathcal{D}=$ & $i\dsum\limits_{\mu =1}^{4}\gamma ^{\mu }\sin ap_{\mu
}-i\cos ap_{0}\left( \dsum\limits_{\mu =1}^{4}\gamma ^{\mu }\upsilon _{\mu
}\right) -i\dsum\limits_{\mu =1}^{4}\gamma ^{\mu }\cos ap_{\mu }$ \\ 
& $+\frac{i}{2}\left( \dsum\limits_{\mu =1}^{4}\gamma ^{\mu }\upsilon _{\mu
}\right) \left( \dsum\limits_{\nu =1}^{4}\upsilon ^{\nu }\cos ap_{\nu
}\right) .$%
\end{tabular}%
\end{equation}%
Using (\ref{A3}), we obtain%
\begin{equation}
\begin{tabular}{ll}
$\mathcal{D}=$ & $\frac{i}{a}\dsum\limits_{\mu =1}^{4}\gamma ^{\mu }\sin
ap_{\mu }-\frac{2i}{a}\Gamma \cos ap_{0}$ \\ 
& $+\frac{i}{a}\Gamma \dsum\limits_{\nu =1}^{4}\cos ap_{\nu }-\frac{i}{a}%
\dsum\limits_{\mu =1}^{4}\gamma ^{\mu }\cos ap_{\mu }.$%
\end{tabular}%
\end{equation}%
But this operator is precisely the one given by (\ref{EG}) with $\Gamma
_{5}=-2i\Gamma $ and $ap_{5}=ap_{0}$.\newline
Comparing with (\ref{A1}), we discover that we should have%
\begin{equation}
\begin{tabular}{llll}
$\cos ap_{0}=1$ & $\Longrightarrow $ & $p_{0}=0$,$\ \ \func{mod}\left( \frac{%
2\pi }{a}\right) $ & .%
\end{tabular}%
\end{equation}%
Let us discuss the meaning of these constraint equations. In our proposal,
the components $\left( p_{0},p_{\mu }\right) $ form a 5-dimensional vector $%
p_{l}$ given by 
\begin{equation}
\begin{tabular}{llll}
$p_{l}=\dsum\limits_{\mu =1}^{4}k_{\mu }.\lambda _{l}^{\mu }$ & , & $%
l=0,...,4$ & ,%
\end{tabular}%
\end{equation}%
where $\mathbf{\lambda }_{l}^{\mu }$ are given by eqs(\ref{US}-\ref{AL}).
Because $\left( \lambda _{0}^{\mu }+\lambda _{1}^{\mu }+\lambda _{2}^{\mu
}+\lambda _{3}^{\mu }+\lambda _{4}^{\mu }\right) =0$, these momenta satisfy
the conservation law%
\begin{equation}
p_{0}=-\left( p_{1}+p_{2}+p_{3}+p_{4}\right) .  \label{P}
\end{equation}%
Now using the fact that the zero modes of the Dirac operator of the \emph{BC}
fermions are either $p_{1}=p_{2}=p_{3}=p_{4}=0$ or $p_{1}=p_{2}=p_{3}=p_{4}=%
\frac{\pi }{a}$, one finds indeed that $p_{0}=0$, $\ \func{mod}\frac{2\pi }{a%
}.$

\subsection{\emph{KW} fermions}

In the \emph{KW} fermions, the Dirac operator $\mathcal{D}_{KW}$ is given by
eq(\ref{n}). By dropping the bare mass $m_{0}$, it reads, up to a global
scale factor, as follows%
\begin{equation}
\begin{tabular}{lllll}
$\mathcal{D}_{KW}$ & $\sim $ &  & $\frac{i}{a}\dsum\limits_{\mu
=1}^{4}\gamma ^{\mu }\sin ak_{\mu }-\frac{i}{a}\gamma _{4}\dsum\limits_{\mu
=1}^{3}\cos ak_{\mu }+\frac{3}{a}i\gamma _{4}$ & .%
\end{tabular}
\label{DK}
\end{equation}%
To make contact between this $4\times 4$ matrix operator and the $\mathcal{D}
$ one following from our proposal, we should solve the equation $\mathcal{D}%
_{KW}=\mathcal{D}$ with \textrm{,}%
\begin{equation}
\begin{tabular}{ll}
$\mathcal{D}\text{ }\sim \text{ }$ & $\frac{1}{a}\dsum\limits_{\mu
=1}^{4}\gamma ^{\mu }\left[ \Omega _{\mu }^{0}-\bar{\Omega}_{\mu }^{0}\right]
\sin ap_{0}+\frac{1}{a}\dsum\limits_{\mu ,\nu =1}^{4}\gamma ^{\mu }\left[
\Omega _{\mu }^{\nu }-\bar{\Omega}_{\mu }^{\nu }\right] \sin ap_{\nu }$ \\ 
& $-\frac{i}{a}\dsum\limits_{\mu =1}^{4}\gamma ^{\mu }\left[ \Omega _{\mu
}^{0}+\bar{\Omega}_{\mu }^{0}\right] \cos ap_{0}-\frac{i}{a}%
\dsum\limits_{\mu ,\nu =1}^{4}\gamma ^{\mu }\left[ \Omega _{\mu }^{\nu }+%
\bar{\Omega}_{\mu }^{\nu }\right] \cos ap_{\nu }.$%
\end{tabular}
\label{DKW}
\end{equation}%
By choosing the $\Omega _{\mu }^{l}$ components as follows,%
\begin{equation}
\Omega _{\mu }^{l}\neq \bar{\Omega}_{\mu }^{l}
\end{equation}%
with%
\begin{equation}
\begin{tabular}{lllll}
$\Omega _{\mu }^{0}=\zeta _{\mu }$ & , & $\Omega _{\mu }^{\nu }=\frac{i}{2}%
\delta _{\mu }^{\nu }+\frac{1}{2}\Theta _{\mu }^{\nu }$ & , & 
\end{tabular}%
\end{equation}%
or equivalently%
\begin{equation}
\begin{tabular}{ll}
$\Omega _{\mu }^{l}=\frac{1}{2}\left( 
\begin{array}{ccccc}
2\zeta _{1} & i+\Theta _{1}^{1} & \Theta _{2}^{1} & \Theta _{3}^{1} & \Theta
_{4}^{1} \\ 
2\zeta _{2} & \Theta _{1}^{2} & i+\Theta _{2}^{2} & \Theta _{3}^{2} & \Theta
_{4}^{2} \\ 
2\zeta _{3} & \Theta _{1}^{3} & \Theta _{2}^{3} & i+\Theta _{3}^{3} & \Theta
_{4}^{3} \\ 
2\zeta _{4} & \Theta _{1}^{4} & \Theta _{2}^{4} & i+\Theta _{3}^{4} & \Theta
_{4}^{4}%
\end{array}%
\right) $ & , \\ 
& 
\end{tabular}%
\end{equation}%
where $\zeta _{\mu }$ and $\Theta _{\mu }^{\nu }$ real. Putting back into (%
\ref{DKW}), we get the reduced operator,%
\begin{equation}
\begin{tabular}{ll}
$\mathcal{D}\text{ }\sim \text{ }$ & $\frac{i}{a}\dsum\limits_{\mu
=1}^{4}\gamma ^{\mu }\sin ap_{\mu }-\frac{2i}{a}\cos ap_{0}\left(
\dsum\limits_{\mu =1}^{4}\gamma ^{\mu }\zeta _{\mu }\right) $ \\ 
& $-\frac{i}{a}\dsum\limits_{\nu =1}^{4}\left( \dsum\limits_{\mu
=1}^{4}\gamma ^{\mu }\Theta _{\mu }^{\nu }\right) \cos ap_{\nu }.$%
\end{tabular}
\label{EE}
\end{equation}%
Comparing with (\ref{DK}), it follows that the choice%
\begin{equation}
\begin{tabular}{llll}
$\zeta _{\mu }=\left( 
\begin{array}{c}
0 \\ 
0 \\ 
0 \\ 
c_{0}%
\end{array}%
\right) $ & , & $\Theta _{\mu }^{\nu }=\left( 
\begin{array}{cccc}
0 & 0 & 0 & 0 \\ 
0 & 0 & 0 & 0 \\ 
0 & 0 & 0 & 0 \\ 
c_{1} & c_{2} & c_{3} & c_{4}%
\end{array}%
\right) $ & ,%
\end{tabular}%
\end{equation}%
leads to%
\begin{equation}
\begin{tabular}{ll}
$\mathcal{D}\text{ }\sim \text{ }$ & $\frac{i}{a}\dsum\limits_{\mu =1}^{4}%
\mathrm{\gamma }_{\mu }\sin ap_{\mu }-\mathrm{\gamma }_{4}\left[ \frac{2i}{a}%
c_{0}\cos ap_{0}\right] $ \\ 
& $-\mathrm{\gamma }_{4}\left( \frac{i}{a}\dsum\limits_{\mu =1}^{4}c_{\mu
}\cos ap_{\mu }\right) .$%
\end{tabular}%
\end{equation}%
By requiring the KW zeros%
\begin{equation}
\begin{tabular}{lllll}
$\left( 1\right) :$ & $p_{1}=p_{2}=p_{3}=0$ & $,$ & $p_{4}=p_{0}=0\ $ & , \\ 
$\left( 2\right) :$ & $p_{1}=p_{2}=p_{3}=0$ & , & $p_{4}=p_{0}=\pi $ & ,%
\end{tabular}%
\end{equation}%
obeying $\sum_{l=0}^{4}p_{l}=0$ $\func{mod}\left( 2\pi \right) $, we obtain
the following constraints on the coefficients%
\begin{equation}
\begin{tabular}{lll}
$\left( 1\right) :$ & $c_{1}+c_{2}+c_{3}+c_{4}+2c_{0}=0$ & $,$ \\ 
$\left( 2\right) :$ & $c_{1}+c_{2}+c_{3}-c_{4}-2c_{0}=0$ & .%
\end{tabular}%
\end{equation}%
A particular solution of these relations is given by%
\begin{equation}
\begin{tabular}{lllll}
$c_{1}=c_{2}=c_{0}$ & $,$ & $c_{3}=c_{4}=-2c_{0}$ & , & $c_{0}=1$.%
\end{tabular}%
\end{equation}

\section{Conclusion and comments}

Motivated by studies on lattice \emph{QCD} and using results on \emph{4D}
graphene, we have developed in this paper a class of \emph{4D} lattice
fermions that live on the hyperdiamond $\mathbb{L}_{4}$ with typical action
given by (\ref{PR}). This lattice action involves a complex tensor $\mathrm{%
\Omega }_{\mu }^{l}$ that captures basic properties on hyperdiamond
fermions; in particular:\newline
(\textbf{1}) it links the \emph{4D} Euclidean vector $\mu $ and the
5-dimensional $\lambda _{l}$ allowing to exhibit explicit both the
underlying $SO\left( 4\right) $ and $SU\left( 5\right) $ symmetries of the
lattice action, \newline
(\textbf{2}) it encodes the data on the zero modes of the Dirac operator $%
\mathcal{D}$. The constraint relation giving the zero modes of the Dirac
operator reads, in terms of the wave vector $\mathbf{K}$, as follows:%
\begin{equation}
\func{Re}\left( \dsum\limits_{l=0}^{4}\mathrm{\Omega }_{\mu }^{l}e^{ia%
\mathbf{K.\lambda }_{l}}\right) =0
\end{equation}%
and its solutions depend indeed on $\mathrm{\Omega }_{\mu }^{l}$ and the
weight vectors $\lambda _{l}$,\newline
(\textbf{3}) it contains as particular examples the \emph{BC} and \emph{KW}
fermions; the corresponding $\left( \mathrm{\Omega }_{\mu }^{l}\right) _{BC}$
and $\left( \mathrm{\Omega }_{\mu }^{l}\right) _{KW}$ were given in the
introduction.\newline
We end this study by noting that one can use the results developed in this
paper to investigate the renormalization properties of the gauged version of
(\ref{PR}) along the same lines as done in \textrm{\cite{5,6}}. For
instance, the free quark propagator following from our proposal is given by 
\begin{equation}
\boldsymbol{S}\left( p\right) \sim a\frac{\gamma ^{\mu }\left( D_{\mu }-\bar{%
D}_{\mu }\right) +aM_{0}}{D^{2}-\bar{D}^{2}+\left( aM_{0}\right) ^{2}},
\end{equation}%
with bare mass $M_{0}$; and 
\begin{equation}
\begin{tabular}{llll}
$D_{\mu }=\dsum\limits_{l=0}^{4}\mathrm{\Omega }_{\mu }^{l}e^{ia\mathbf{p}.%
\mathbf{\lambda }_{l}}$ & , & $\bar{D}_{\mu }=\dsum\limits_{l=0}^{4}\mathrm{%
\bar{\Omega}}_{\mu }^{l}e^{-ia\mathbf{p}.\mathbf{\lambda }_{l}}$ & ,%
\end{tabular}%
\end{equation}%
and where $D^{2}=D_{\mu }D^{\mu }$, $\bar{D}^{2}=\bar{D}_{\mu }\bar{D}^{\mu
} $. The gauging of (\ref{PR}) may be done by making $\mathrm{\Omega }_{\mu
}^{l}$ a local field; i.e by allowing it to depend on the position. Progress
in this direction will be given in a future occasion.

\begin{acknowledgement}
the authors would like to thank M.Bousmina for discussions; Tatsuhiro Misumi
and Simon Catterall for drawing our attention to their works \cite{14,15,16}%
. L.B Drissi would like to thank ICTP for associate membership and E.H Saidi
thanks URAC-09, CNRST.
\end{acknowledgement}


\begin{thebibliography}{99}
\bibitem{1} M. Creutz, JHEP 0804 $\left( 2008\right) $ 017,
[arXiv:0712.1201],

\bibitem{2} A. H.Castro-Neto et al. Rev. Mod. Phys.81, 109 $\left(
2009\right) $,

\bibitem{3} L. B Drissi, E.H Saidi, M.Bousmina, Nucl Phys B, Vol 829,
p.523-533.

\bibitem{4} A. Borici, Phys. Rev. D78 $\left( 2008\right) $ 074504,
[arXiv:0712.4401],

\bibitem{5} S. Capitani, J. Weber, H. Wittig, Phys. Lett. B 681, 2009,
105-112, arXiv:0907.2825

\bibitem{6} S. Capitani, M. Creutz, J. Weber, H.Wittig, JHEP 1009:027,2010,
arXiv:1006.2009.

\bibitem{7} L.H. Karsten, Phys. Lett. B104 (1981) 315,

\bibitem{8} F. Wilczek, Phys. Rev. Lett. 59 (1987) 2397.

\bibitem{9} L.B Drissi, E.H Saidi, M. Bousmina,\emph{\ }J. Math. Phys. 52,
022306 (2011)

\bibitem{10} L.B Drissi, E.H Saidi, M. Bousmina,\emph{\ Four Dimensional
Graphene}, CPM-11-01, Phys. Rev. D (2011), in press.

\bibitem{11} P. F. Bedaque, M. I. Buchoff, B. C. Tiburzi, A. Walker-Loud,
Phys. Rev. D78 $\left( 2008\right) $ 017502, [arXiv:0804.1145],

\bibitem{12} P. F. Bedaque, M. I. Buchoff, B. C. Tiburzi, A. Walker-Loud,
Phys. Lett. B662 $\left( 2008\right) $ 449, [arXiv:0801.3361],

\bibitem{13} Taro Kimura, Tatsuhiro Misumi, Prog.Theor.Phys.123: 63-78, $%
\left( 2010\right) $, arXiv:0907.3774,

\bibitem{14} Michael Creutz, Tatsuhiro Misumi, \emph{Classification of
Minimally Doubled Fermions}, Phys.Rev.D82:074502,2010, arXiv:1007.3328,

\bibitem{15} Simon Catterall, David B. Kaplan, Mithat Unsal, \emph{Exact
lattice supersymmetry}, arXiv:0903.4881, To be published in Physics Reports,

\bibitem{16} Simon Catterall, Eric Dzienkowski, Joel Giedt, Anosh Joseph,
Robert Wells, \emph{Perturbative renormalization of lattice N=4 super
Yang-Mills theory}, arXiv:1102.1725.
\end{thebibliography}
\end{document}